\documentclass[12pt,letterpaper]{article}
\usepackage{stmaryrd}
\usepackage{verbatim}
\usepackage{float}
\usepackage{booktabs}
\usepackage{amsmath}
\usepackage{amssymb}
\usepackage{graphicx}
\usepackage{setspace}
\usepackage{natbib}
\usepackage{algorithm}
\usepackage{algorithmic}
\usepackage{xcolor, colortbl,epstopdf}
\usepackage{amsthm}
\usepackage{rotating}
\usepackage{stmaryrd}
\newtheorem{lemma}{Lemma}

\definecolor{lightgray}{gray}{0.9}
\newtheorem{prop}{Proposition}
\bibpunct{(}{)}{;}{a}{,}{,}

\newcommand{\ba}{\mathbf{a}}
\newcommand{\bA}{\mathbf{A}}
\newcommand{\bfb}{\mathbf{b}}
\newcommand{\bB}{\mathbf{B}}

\newcommand{\bD}{\mathbf{D}}
\newcommand{\bE}{\mathbf{E}}
\newcommand{\be}{\mathbf{e}}

\newcommand{\bg}{\mathbf{g}}
\newcommand{\bG}{\mathbf{G}}

\newcommand{\bI}{\mathbf{I}}
\newcommand{\bK}{\mathbf{K}}

\newcommand{\bm}{\mathbf{m}}
\newcommand{\bM}{\mathbf{M}}

\newcommand{\bP}{\mathbf{P}}

\newcommand{\bS}{\mathbf{S}}
\newcommand{\bu}{\mathbf{u}}
\newcommand{\bU}{\mathbf{U}}
\newcommand{\bv}{\mathbf{v}}

\newcommand{\bx}{\mathbf{x}}
\newcommand{\bX}{\mathbf{X}}
\newcommand{\by}{\mathbf{y}}
\newcommand{\bY}{\mathbf{Y}}
\newcommand{\bV}{\mathbf{V}}

\newcommand{\bZ}{\mathbf{Z}}

\newcommand{\mA}{\mathcal{A}}
\newcommand{\mB}{\mathcal{B}}

\newcommand{\mE}{\mathcal{E}}

\newcommand{\mG}{\mathcal{G}}

\newcommand{\mM}{\mathcal{M}}

\newcommand{\mX}{\mathcal{X}}
\newcommand{\mY}{\mathcal{Y}}

\newcommand{\vect}[1]{\boldsymbol #1}

\newcommand{\vOmega}{\vect{\Omega}}
\newcommand{\vomega}{\vect{\omega}}
\newcommand{\vSigma}{\vect{\Sigma}}

\newcommand{\tvec}{\text{vec}}

\renewcommand{\epsilon}{\varepsilon}
\renewcommand{\hat}{\widehat}
\renewcommand{\tilde}{\widetilde}
\renewcommand{\leq}{\leqslant}
\renewcommand{\geq}{\geqslant}

\newcommand{\distn}[1]{\mathcal{#1}}

\author{Yaling Qi\thanks{I am particularly grateful to Joshua Chan for his invaluable guidance and support, and to Mario Crucini, Yong Bao, Justin Tobias, and Mohitosh Kejriwal for helpful comments and discussions.} \\ Purdue University }

\begin{document}

\title{Large Bayesian Tensor Autoregressions}


\maketitle



\thispagestyle{empty}

\begin{abstract}
The availability of multidimensional economic datasets has grown significantly in recent years. An example is bilateral trade values  across goods among countries, comprising three dimensions—importing countries, exporting countries, and goods—forming a time series of three-dimensional arrays, called tensor time series. This paper introduces a general Bayesian tensor autoregressive framework to analyze the dynamics of large, multidimensional time series with a particular focus on international trade across different countries and sectors. Departing from the standard homoskedastic assumption in this literature, we incorporate flexible stochastic volatility into the tensor autoregressive models. The proposed models could capture time-varying volatility due to the COVID-19 pandemic and recent outbreaks of war. To address computational challenges and mitigate overfitting, we develop an efficient sampling method based on low-rank Tucker decomposition and hierarchical shrinkage priors. Additionally, we provide an innovative factor interpretation of the model showing how the Tucker decomposition projects large-dimensional disaggregated trade flows onto global driving factors. 




\end{abstract}
\newpage
\section{Introduction}
 Recent years have seen the proliferation of big data across biological, physical, economic, and financial fields, which provides unprecedented access to diverse  multidimensional datasets, including, but not limited to, increasingly structured and complex time-series data.  
 Some high-dimensional time-series data have a natural tensor representation. For example, in the global economy, bilateral trade flows disaggregated by commodity categories can be represented as a three-dimensional array, which we treat as a time series with tensor values, with each mode corresponding to importing countries, exporting countries, and goods, respectively. 
The availability of multiway trade data motivates two related research questions. First, which countries and commodity categories contribute most to the common dynamics of global trade, as opposed to idiosyncratic bilateral fluctuations? Second, what are the major latent global factors driving the cross-country, cross-commodity trade tensor? Answering these questions provides additional insight for identifying economically meaningful global shocks.

Early studies tend to directly stack the high-dimensional data into a long vector and apply vector autoregressions (VAR) or factor models to analyze the large data. However, these approaches have some limitations. First, this transformation disregards the inherent structural information within the datasets. In the trade example, we can expect cross-sectional dependencies between different countries and goods. However, vectorization, by mixing rows and columns, obscures the underlying multiway dependence structure.  Although previous research has extended the standard VAR framework to account for interdependencies between countries—for instance, through panel VAR \citep{canova2013panel} or global VAR \citep{pesaran2004modeling, dees2007exploring, bussiere2009modelling}
, they are limited to aggregate trade values at the country level, thereby overlooking the disaggregated multilateral and multi-category trade distribution across countries. Moreover, analyzing such networks involves handling hundreds or even tens of thousands of time series, making traditional VAR models extremely time-consuming or even infeasible due to their substantial computational complexity \citep{banbura2010large, koop13, CCM15}.

Recent studies have started to explore multidimensional economic datasets while preserving the tensor structure. Many of them focus on matrix-variate regressions, including both autoregressive and factor models \citep{Hoff15,ding2018matrix, CXY21, chan2025large, wang2019factor, chen2020constrained,zhang2024bayesian}.  A prominent class of matrix autoregressive models assumes a bilinear structure with two coefficient matrices, where the two transition matrices are typically interpreted as capturing either column-wise or row-wise interactions, depending on the data. These studies provide deeper insights into cross-sectional dependencies or spillovers. 

By contrast, third-order tensor regressive models remain relatively scarce, particularly in the context of large-scale tensor data. A major concern in tensor regressive models is the proliferation of parameters in high-dimensional settings, often referred to as the “curse of dimensionality,” which arises from the large number of parameters these models entail. This leads to two major estimation problems. On the one hand, if the number of unknown parameters far exceeds the sample size, accurate estimation becomes very difficult without regularization. Previous Bayesian studies have often employed hierarchical shrinkage priors to address this issue, which can improve estimation and inference by shrinking negligible coefficients toward zero while preserving relatively large coefficients. On the other hand, estimating a large number of free parameters significantly increases computational complexity. The estimation process even becomes particularly infeasible when handling datasets comprising thousands of time series.

To address this challenge, tensor decomposition provides a widely used dimension-reduction approach for high-dimensional data by exploiting the assumption that the underlying tensor has low multilinear rank. This method extends the concept of Singular Value Decomposition (SVD) from matrices to higher-order tensors, enabling more efficient computation. Two widely used tensor decomposition methods are the PARAFAC/CANDECOMP (CP) decomposition \citep{harshman1970foundations, carroll1970analysis} and the Tucker decomposition \citep{tucker1963implications, levin1965three, tucker1966some}. CP decomposition breaks down the multidimensional coefficient tensor into a sum of rank-one tensors, each constructed as the outer product of marginal vectors along each dimension. This approach reduces millions of elements in the coefficient tensor to a manageable number for efficient and reliable estimation.  Tucker decomposition extends the CP decomposition by representing a tensor as a smaller core tensor multiplied by "loading" matrices along each mode. Compared to CP decomposition, Tucker decomposition is more flexible as it allows for different ranks across dimensions and capture more complicate associations among marginals.  
Particularly, CP decomposition can be viewed as a special case of Tucker decomposition (see Section \ref{Deco}).

  
Most of the existing tensor regression literature is designed for settings in which either the predictors or the responses are tensor-valued, but not both, such as scalar-on-tensor, vector-on-tensor, and their reverse settings \citep{zhou2013tensor, zhou2015bayesian, rabusseau2016low, guhaniyogi2017bayesian, li2017parsimonious, guhaniyogi2021bayesian}. \cite{zhou2013tensor} recognize the limitations of traditional methods in analyzing complex, multidimensional medical imaging data and propose generalized linear regressive models to accommodate tensor-valued imaging data as covariates, and extend their framework by modeling the coefficient tensor using Tucker decomposition in \cite{li2018tucker}. However, traditional regularization approaches used in the frequentist literature, such as Lasso and Ridge, have a limitation in that they rely heavily on manually selected tuning parameters, which are sensitive to the model specification, such as the tensor dimension. To address this, \cite{guhaniyogi2017bayesian} develop a Bayesian framework incorporating a novel class of multiway shrinkage priors, which automatically induce sparsity and enable reliable identification of important coefficients. Subsequently, \cite{spencer2020joint} extend this framework by introducing rank-specific prior scale parameters through a stick-breaking construction, allowing for efficient shrinkage across ranks of CP decomposition.          

To date, the literature on tensor-on-tensor regression has been developing but remains limited \citep{Hoff15, lock2018tensor,wang2024high,wang2024bayesian}, particularly within the Bayesian framework. 
\cite{Hoff15} introduces a general multilinear tensor regressive framework to analyze social networks in the context of tensor-valued longitudinal and multivariate relational data. More recently, \cite{billio2023bayesian} proposed a tensor-on-(vectorized) tensor setting in Bayesian framework with CP decomposition, and applied it to a two-layer network of trade and credit linkages among ten countries. 
The tensor regression in \cite{wang2024bayesian} allows for rich dependence within each tensor observation but remains static, as the likelihood factorizes across observations and temporal dependence is not explicitly modeled. 
Additionally, as for the frequentist literature, \cite{wang2024high} introduce both convex and non-convex estimation methods for tensor autoregressions based on the Tucker decomposition and apply them to trade networks. They also mention an advantage of the Tucker decomposition over the CP decomposition, namely that it facilitates dynamic factor interpretations of the multilinear model. However, their analysis is conducted under the assumption of homoskedastic errors and does not provide a framework for interpreting the resulting factor time series in terms of their underlying economic content.


Building on the existing literature, we develop a novel and more general Bayesian tensor-on-tensor autoregressive framework based on the Tucker decomposition,
and provide meaningful complement to prior work. Our paper advances the development of tensor autoregressions along three key dimensions, spanning both methodological and empirical contributions. 

First, as a growing body of literature emphasizes the importance of stochastic volatility for estimation and inference in macroeconomic models \citep{CS05,SZ06,KK13,CCMM22,chan2023comparing}—particularly during periods of heightened uncertainty—we incorporate stochastic volatility to capture time-varying economic uncertainty. We also utilize a specialized hierarchical shrinkage prior—an adapted version of the multiway stick-breaking shrinkage prior proposed in \cite{guhaniyogi2020bayesian} for the CP decomposition—to automatically induce sparsity, even across ranks, and enhance the reliability of the estimation.

Regarding the second contribution, existing studies primarily focus on small- or medium-sized tensor data \citep{guhaniyogi2017bayesian, billio2023bayesian}. For example, \cite{billio2023bayesian} analyze bilateral trade and credit flows among 10 countries with $(\mY_t)$ of dimension $(10 \times 10 \times 2)$, totaling 200 time series.  In contrast, we develop a computationally efficient Bayesian estimation approach that scales to large datasets through a more parsimonious parameterization. Rather than estimating the fourth-order marginal of length $I_1I_2I_3$ directly, we decompose it into three mode-specific marginals of dimensions $I_1$, $I_2$, and $I_3$, thereby substantially reducing the number of parameters.
 Our framework is particularly well suited to high-dimensional settings; for instance, a third-order tensor $(\mY_t)$ of size ($18 \times 18 \times 15$) contains 4,860 elements.  Moreover, we extend the commonly used CP decomposition to the more flexible Tucker decomposition, allowing for heterogeneous ranks across modes and introducing a core tensor to capture complex interactions among them. 

The final contribution lies in uncovering the economic content of tensor autoregressive models. We interpret the Tucker model as projecting high-dimensional, disaggregated time series onto a lower-dimensional factor space through mode-specific loading matrices. 
This factor structure provides economic content to the Tucker decomposition and implicitly addresses the research questions posed earlier. In particular, the estimated loading matrices identify a small number of systemically important countries—such as the United States, Germany, and France—as key drivers of global trade dynamics, together with economically meaningful clustering patterns across commodity groups. Moreover, the common factors extracted from the large-scale disaggregated trade tensor are strongly aligned with several key global macroeconomic indicators, most notably the Brent crude oil price index and the Real Broad Dollar index. This finding suggests that the latent trade factors capture important global economic conditions and supports their interpretation as common drivers of global trade fluctuations.

 %


The rest of the paper is structured as follows: the next section introduces the flexible framework for autoregressive tensor models. Section 3 provides details of posterior computation under Tucker decomposition. Section 4 presents Monte Carlo simulation results comparing various competing methods under a range of data-generating processes. Section 5 applies the proposed efficient estimation method to analyze the multi-category international bilateral trade flows. Finally, section 6 concludes the paper and discusses potential directions for further research.

\section{A Flexible Framework for Tensor Autoregressive Models}

In this section, we propose a general Bayesian tensor autoregressive framework with stochastic volatility and introduce tensor decomposition as a widely used dimension-reduction technique. The resulting framework flexibly accommodates a wide range of time-varying error structures and enables efficient modeling of high-dimensional time series.
\subsection{Notations}
 We begin by introducing the basic notation and algebra used throughout the paper, with a particular focus on tensors. Additional details can be found in \citet{kolda2009tensor}. In general, a tensor is a multidimensional array. A first-order tensor is a vector (denoted by boldface lowercase letters, e.g., $\bx$); a second-order tensor is a matrix (denoted by boldface uppercase letters, e.g., $\bX$); and tensors of order three or higher are referred to as higher-order tensors (denoted by boldface Euler script letters, e.g., $\mX$).

The Kronecker product is another fundamental operation used in this paper.  Let $\mathbf{A}$ be an $m \times n$ matrix and $\mathbf{B}$ be a $p \times q$ matrix. The Kronecker product $\mathbf{A} \otimes \mathbf{B}$ is defined as a $pm \times qn$ block matrix obtained by multiplying each element of $\mathbf{A}$ by the entire matrix $\mathbf{B}$, that is,
$$\mathbf{A} \otimes \mathbf{B}=\left[\begin{array}{ccc}a_{11} \mathbf{B} & \cdots & a_{1 n} \mathbf{B} \\ \vdots & \ddots & \vdots \\ a_{m 1} \mathbf{B} & \cdots & a_{m n} \mathbf{B}\end{array}\right].$$

Next, we introduce the vectorization operator. The vectorization of a tensor $\mX \in \mathbb{R}^{I_1 \times I_2 \times \cdots \times I_N}$ stacks the $N$-dimensional array into a long vector $\bx \in \mathbb{R}^{\prod_{i=1}^N I_i}$. Specifically, the $(i_1, \ldots, i_N)^{th}$ element of $\mX$ corresponds to the $i^{th}$ element of $\bx$, where
$i = 1 + \sum_{k=1}^N (i_k - 1) d_k, \quad \text{with } d_k = \prod_{m=1}^{k-1} I_m.$

Another important operation is mode-n matricization. The mode-$n$ matricization of a tensor $\mX \in \mathbb{R}^{I_1 \times I_2 \times \cdots \times I_N}$ is denoted by $\mX_{(n)}$ with size $I_n \times \prod_{i\neq n} I_i$. It arranges the mode-$n$ fibers to be the columns of the resulting matrix.  Through mode-$n$ matricization, the tensor element $\left(i_1, i_2, \ldots, i_N\right)$ maps to matrix element $\left(i_n, j\right)$, where               
$
j=1+\sum_{\substack{\ k \neq n}} \left(i_k-1\right) J_k$, with $J_k=\prod_{\substack{ m \neq n}}^{k-1} I_m
$.


Finally, we define the generalized inner product. The generalized inner product of two different-sized tensors $\mX \in \mathbb{R}^{I_1 \times \cdots \times I_n \times \cdots \times I_N}$ and $\mY  \in \mathbb{R}^{I_n \times \cdots \times I_N}$ is the sum of the products of their corresponding entries, resulting in an $(n-1)^{th}$-order tensor of dimension $I_1 \times \cdots \times I_{n-1}$, of which $(i_1, \cdots, i_{n-1})^{th}$ element is given by:
$$
\langle\mX, \mY\rangle _{i_1, \cdots,i_{n-1}}=\sum_{i_n=1}^{I_n} \cdots \sum_{i_N=1}^{I_N} \mX_{i_1, \cdots,i_{n-1}, i_n,\cdots, i_N} \mY_{i_n, \cdots, i_N} .
$$



\subsection{Introduction to Tensor Autoregressive Models }

Consider a third-order tensor time series $\mY_t \in \mathbb{R}^{I_1 \times I_2 \times I_3}$ for $t = 1, \dots, T$, evolving according to an autoregressive process. The dynamics of $\mY_t$ are specified by a tensor autoregressive model of order $p$, denoted TAR($p$):
\begin{equation}\label{BTAR}
\mY_t = \mA_t + \sum_{j=1}^p \langle \mB_j, \mY_{t-j} \rangle + \mE_t,
\end{equation}
where $\mY_{t}$, $\mA_t$, and $\mE_t$ are all third-order tensors of dimension $I_1 \times I_2 \times I_3$. The term $\mA_t$ captures deterministic components, such as intercepts or time trends; for example, a linear trend specification can be written as $\mA_t = \mA_0 + \mA_1 t$. 

In addition, $\mB_j \in \mathbb{R}^{I_1 \times I_2 \times I_3 \times I_1 \times I_2 \times I_3}$ denote sixth-order coefficient tensors conformable with the generalized inner product $\langle \cdot, \cdot \rangle$.
Elementwise, let $\mY_{t,i_1,i_2,i_3}$ ($\mE_{t,i_1,i_2,i_3}$) denote the $(i_1,i_2,i_3)^{th}$ entry of $\mY_t$ ($\mE_t$); then we have

\begin{equation}\mY_{t,i_1,i_2,i_3}=\mA_{t,i_1,i_2,i_3}+\sum_{j=1}^p\sum_{i_4=1}^{I_1}\sum_{i_5=1}^{I_2}\sum_{i_6=1}^{I_3}\mB_{j,i_1,i_2,i_3,i_4,i_5,i_6}\mY_{t-j,i_4,i_5,i_6}+\mE_{t,i_1,i_2,i_3}\end{equation}
The lagged effect of $\mY_{t-j,i_4,i_5,i_6}$  on $\mY_{t,i_1,i_2,i_3}$ is captured by $\mB_{j,i_1,i_2,i_3,i_4,i_5,i_6}$.  If we want to study the dynamics of bilateral trade among \( n \) countries across \( m \)  commodity categories, we can represent the data as an \( n \times n \times m \) tensor-valued \(\mY_t\), where $\mY_{t,i_1,i_2,i_3}$ denotes the import value of  country \( i_1 \) from the exporting country \( i_2 \) for commodity category \( i_3 \) at time \( t \). The three dimensions correspond to importers, exporters, and commodity categories, respectively.

 In contrast to \citet{billio2023bayesian}, who stack tensor-valued predictors at each time period into a single, long vector for inference, we preserve the inherent multiway structure of both predictors and responses.  Accordingly, the autoregressive coefficients in the proposed model are represented by sixth-order tensors $\mB_j$, rather than fourth-order tensors in \citet{billio2023bayesian}, of dimension $I_1 \times I_2 \times I_3 \times I$, where $I = I_1I_2I_3$. This representation exploits the multiway structure of the data and leads to a substantially more parsimonious parameterization.


\subsection{Stochastic Volatility}\label{SV}

Previous studies have demonstrated the importance of modeling time-varying volatility in macroeconomic models \citep{KK13, CCM16, CCM19, chan2020composite, chan2023comparing}, particularly during periods of increased uncertainty, such as the global financial crisis, the COVID-19 pandemic, and more recent episodes of escalating regional geopolitical tensions.
 Building on the framework of \cite{chan2025large}, we develop a generalized specification that can accommodate multiple forms of stochastic volatility through a latent mixture parameter, $\omega_t$. 
 
 The error term $\mE_t$ given the latent variable $\omega_t$ follows a conditional Gaussian distribution with a Kronecker product structure:: 
\begin{equation} \label{error}
\tvec(\mE_t) \sim \mathcal{N}(\mathbf{0}_I, \omega_t\vSigma),
\text{ where } \vSigma =\vSigma_{3}\otimes\vSigma_{2}\otimes\vSigma_{1}
\end{equation}
where $\tvec(\mE_t) = \tvec(\bE_{t,(1)})$. $\bE_{t,(1)}$ is the mode-1 matricization of $\mE_t$, and $\tvec(\bE_{t,(1)})$ denotes the operation that stacks the columns of the matrix $\bE_{t,(1)}$ vertically into a single vector. 
  \(\vSigma_{1}\), \(\vSigma_{2}\), and \(\vSigma_{3}\) are \(I_1 \times I_1\), \(I_2 \times I_2\), and \(I_3 \times I_3\) covariance matrices, respectively. If \(\omega_t = 1\) for all \(t\), 
  the error covariance matrix becomes time-invariant, corresponding to a homoskedastic setting. 
  Next, we introduce two stochastic volatility specifications within this framework.
 
  In the first example, the common stochastic volatility (CSV) model, originally proposed by  \citet{carriero2016common}, assumes that log-volatility, $h_t = \log(\omega_t)$, follows a stationary AR(1) process:
\[
h_t = \phi h_{t-1} + \epsilon_t
\]
where \(\epsilon_t \sim \distn{N}(0, \sigma^2)\), $|\phi|<1$ and the initial condition is defined as $h_1\sim\distn{N}(0,\sigma^2/(1-\phi^2))$. Under this AR(1) specification, volatility at time \(t\) strongly depends on previous periods, leading to more persistent effects of past volatility in the CSV model.

Additionally, we account for potential outliers using the explicit outlier component proposed by \cite{stock2016core}. Specifically, we define \(\omega_t = o_t^2\), where \(o_t\) is either a point mass at 1 or follows a uniform distribution over the interval \((2,10)\). When \(o_t = 1\), the economy is in a normal state, whereas when \(o_t \geq 2\), economic uncertainty increases, which we classify as an outlier status, capturing periods of abnormal volatility or structural disruptions. The sampling algorithm for \(\omega_t\) is detailed in the appendix of \cite{chan2025large}.


 The Kronecker structure is adopted to achieve dimensionality reduction. Without this structure, sampling $\vSigma$ would require estimating $I^2$ free parameters, making estimation both unreliable with limited samples and computationally very expensive with a computational cost of $\mathcal{O}(I^6)$. In contrast, imposing a Kronecker structure drastically reduces the number of free parameters to only $\sum_{i=1}^3{I_i^2}$ and lowers the computational cost to $\mathcal{O}(\max(I_1, I_2, I_3)^6)$, thereby making the estimation more reliable and computationally tractable even in high-dimensional settings.
 
In  addition, the structure has a natural and interpretable decomposition, which can be expressed through its mode-$i$ matricization.  It can be readily shown that the mode-$i$ matricization of $\mE_t$, denoted by $\bE_{t,(i)}$, satisfies the following property:
\begin{equation}\label{eq:E_mat}
\tvec(\bE_{t,(i)})\sim \mathcal{N}(\boldsymbol{0}, \omega_t\tilde{\vSigma}_{i} )
\end{equation}
where $\tilde{\vSigma}_{1}=\vSigma_{3}\otimes\vSigma_{2}\otimes\vSigma_{1}$, $\tilde{\vSigma}_{2}=\vSigma_{3}\otimes\vSigma_{1}\otimes\vSigma_{2} $, and $\tilde{\vSigma}_{3}=\vSigma_{2}\otimes\vSigma_{1}\otimes\vSigma_{3} $. Under the homoskedastic setting, where $\omega_t=1$, the error term has the following equivalent representation:
$$
\bE_{t,(1)} =\vSigma_1^{1/2} \bZ_{1t} (\vSigma_3 \otimes \vSigma_2)^{1/2}, \quad
\bE_{t,(2)} =\vSigma_2^{1/2} \bZ_{2t} (\vSigma_3 \otimes \vSigma_1)^{1/2}, \quad
\bE_{t,(3)} = \vSigma_3^{1/2} \bZ_{3t} (\vSigma_2 \otimes \vSigma_1)^{1/2},
$$
where $\bZ_{1t}$, $\bZ_{2t}$, and $\bZ_{3t}$ are matrices of dimensions $I_1 \times (I_2 I_3)$, $I_2 \times (I_1 I_3)$, and $I_3 \times (I_1 I_2)$, respectively, each consisting of independent standard normal entries. This formulation indicates that $\vSigma_1$, $\vSigma_2$, and $\vSigma_3$ represent the covariance matrices along the first, second, and third tensor modes, respectively. More generally, the covariance between the $(i_1, j_1, k_1)$ and $(i_2, j_2, k_2)$ elements of $\mE_t$ is given by:
$$
\mathrm{cov}(e_{t, i_1, j_1, k_1}, e_{t, i_2, j_2, k_2}) = \sigma_{1, i_1, i_2} \, \sigma_{2, j_1, j_2} \, \sigma_{3, k_1, k_2}
$$


\subsection{Relation to VARs}                                                              
In this section,  The tensor autoregressive (TAR) model can be expressed in the form of a vector autoregressive (VAR) model. Specifically, by defining $\by_t=\tvec(\mY_t)$, $\ba_t=\tvec(\mA_t)$, and $\be_t=\tvec(\mE_t)$, the TAR($p$) model can be equivalently written in vectorized form as
\begin{equation}\label{eq:VAR}
\by_t = \ba_t + \sum_{j=1}^p \hat{\bB}_j \by_{t-j} + \be_t, \quad \be_t \sim \mathcal{N}(\mathbf{0}_I, \omega_t\vSigma)
\end{equation}
where each coefficient matrix $\hat{\bB}_j \in \mathbb{R}^{I \times I}$ is obtained by reshaping the coefficient tensor $\mB_j$, satisfying
$\tvec(\hat{\bB}_j)=\tvec(\mB_j).$ Under this representation, the lagged effect of $\mY_{t-j,i_4,i_5,i_6}$ on $\mY_{t,i_1,i_2,i_3}$ is characterized by the $(k_1,k_2)^{th}$ entry of $\hat{\bB}_j$, where
$
k_1=i_1+(i_2-1)I_1+(i_3-1)I_1I_2,
k_2=i_4+(i_5-1)I_1+(i_6-1)I_1I_2.
$ In the special case of $p=1$, the TAR($1$) model reduces to
\begin{equation}\label{eq:VAR1}
\by_t
=
\ba_t
+
\hat{\bB}_1 \by_{t-1}
+
\be_t,
\qquad
\be_t \sim \mathcal{N}(\mathbf{0}_I,\omega_t\vSigma).
\end{equation}

\subsection{Tensor Decompositions}\label{Deco}

In the rest of the paper, for expositional simplicity, we focus on the tensor autogressions with only one lag, denoted TAR(1) model and simplify the notation by letting \(\mB\) represent \(\mB_1\). 
\begin{equation}\label{BTAR1}
\mY_t = \mA_t + \langle \mB, \mY_{t-1} \rangle + \mE_t, \tvec(\mE_t) \sim \mathcal{N}(\mathbf{0}_I, \omega_t\vSigma_{3}\otimes\vSigma_{2}\otimes\vSigma_{1})
\end{equation}
As noted above, tensor autoregressive models face a key challenge in high-dimensional settings: the rapid growth in the number of parameters, commonly referred to as the “curse of dimensionality.” For example, if \(I_1 = I_2 = 18\) and \(I_3 = 15\), \(I\) would be up to 4,860, resulting in tens of millions (\(23,619,600\)) of parameters to estimate, which is practically impossible. To address this challenge, we employ tensor decomposition, a widely used dimension-reduction technique for high-dimensional tensors. 

Tensor decomposition is a higher-order generalization of matrix singular value decomposition (SVD). Assume that $\bM$ is an $m \times n$ matrix of rank $ R $ with singular value decomposition $( \bU, \bD, \bV )$. Then $\bM$ can be expressed as a sum of $R$ rank-one matrices:
$$
\bM = \bU \bD \bV' = \sum_{\ell=1}^{R} d_{\ell} \bu_{\ell} \bv_{\ell}'
$$
where $\bD\in\mathbb{R}^{R\times R}$ is a diagonal matrix with diagonal entries $d_{\ell}$ for $\ell = 1, \dots, R$, and $\bU\in\mathbb{R}^{m\times R}$ and $\bV\in\mathbb{R}^{n\times R}$ are orthogonal matrices whose columns consist of the left and right singular vectors, $\bu_{\ell}$ and $\bv_{\ell}$, respectively. The term $\bu_{\ell} \bv_{\ell}'$ denotes the outer product of $\bu_{\ell}$ and $\bv_{\ell}$—that is, the $m\times n$ matrix whose $(i,j)^{th}$ entry equals $\bu_{\ell}(i)\bv_{\ell}(j)$. 
Furthermore, the partial sum $ \sum_{\ell=1}^{r} d_{\ell} \bu_{\ell} \bv_{\ell}^{\top} $ provides the best rank-$r$ approximation to the matrix $ \bM $ in the sense of minimizing the Frobenius norm of the approximation error.

The two most widely used tensor decomposition methods$ - $CP decomposition \citep{harshman1970foundations, carroll1970analysis} and Tucker decomposition \citep{tucker1963implications,levin1965three,tucker1966some}, extend the matrix SVD to higher-order arrays. CP decomposition approximates a tensor as a sum of \(R\) rank-one tensors. Each rank-one tensor is constructed as the outer product of mode-specific marginal vectors. For example,
the CP decomposition of a third-order tensor \(\mM \in \mathbb{R}^{I_1 \times I_2 \times I_3}\), 
\[
\mM \approx \sum_{r=1}^{R} \mM_r
= \sum_{r=1}^{R}
\bm_r^{(1)} \circ
\bm_r^{(2)} \circ
\bm_r^{(3)},
\]
where each \(\mM_r \in \mathbb{R}^{I_1 \times I_2 \times I_3}\) is a rank-one tensor, and
\(\bm_r^{(1)} \in \mathbb{R}^{I_1}, \bm_r^{(2)} \in \mathbb{R}^{I_2}\), and \(\bm_r^{(3)} \in \mathbb{R}^{I_3}\), denote the corresponding marginal vectors. In particular, as for the $(i_1,i_2,i_3)^{th}$ element of $\mM$,
$
\mM_{i_1,i_2,i_3}
=
\sum_{r=1}^R
\bm_{i_1,r}^{(1)}
\bm_{i_2,r}^{(2)}
\bm_{i_3,r}^{(3)},$
where $\bm_{i_1,r}^{(1)}$ denotes the $i_1^{th}$ element of the vector $\bm_r^{(1)}$, and similarly for $\bm_{i_2,r}^{(2)}$ and $\bm_{i_3,r}^{(3)}$.




The Tucker decomposition is a generalization of the CP decomposition. It was first introduced by \citet{tucker1963implications} and was later developed in \citet{levin1965three} and \citet{tucker1966some}. 
Tucker decomposition factorizes a tensor into a smaller core tensor and multiple loading matrices along each mode. For a third-order tensor \(\mM \in \mathbb{R}^{I_1 \times I_2 \times I_3}\), the Tucker decomposition can be represented as
\begin{equation}
\mM \approx \llbracket \mG; \bM_1, \bM_2, \bM_3 \rrbracket
= \sum_{r_1=1}^{R_1}\sum_{r_2=1}^{R_2}\sum_{r_3=1}^{R_3}
g_{r_1 r_2 r_3}
\bm_{r_1}^{(1)} \circ
\bm_{r_2}^{(2)} \circ
\bm_{r_3}^{(3)} 
\end{equation}
Elementwise,
$\mM_{i_1,i_2,i_3}
=
\sum_{r_1=1}^{R_1}
\sum_{r_2=1}^{R_2}
\sum_{r_3=1}^{R_3}
g_{r_1 r_2 r_3}
\bm_{i_1,r_1}^{(1)}
\bm_{i_2,r_2}^{(2)}
\bm_{i_3,r_3}^{(3)}$.
Here, the marginal vectors $\bm_{r_1}^{(1)}$, $\bm_{r_2}^{(2)}$, and $\bm_{r_3}^{(3)}$ are of dimensions $I_1$, $I_2$, and $I_3$, respectively,
with $r_i = 1, \dots, R_i$. The integer \(R_i\) denotes the rank for $i^{th}$ dimension. The loading matrices are defined as collections of mode-specific marginal vectors:
$
\bM_i = \left[ \bm_{1}^{(i)}, \dots, \bm_{R_i}^{(i)} \right], i = 1,2,3.
$
The smaller tensor \(\mG \in \mathbb{R}^{R_1 \times R_2 \times R_3}\) is the core tensor, of which the entries quantify interactions among the latent vectors across modes. Typically, we impose \(R_1 \le I_1\), \(R_2 \le I_2\), and \(R_3 \le I_3\), so that the core tensor provides a compressed representation of the original tensor \(\mM\).

\begin{figure}[H]
\centering
\includegraphics[width=0.8\textwidth]{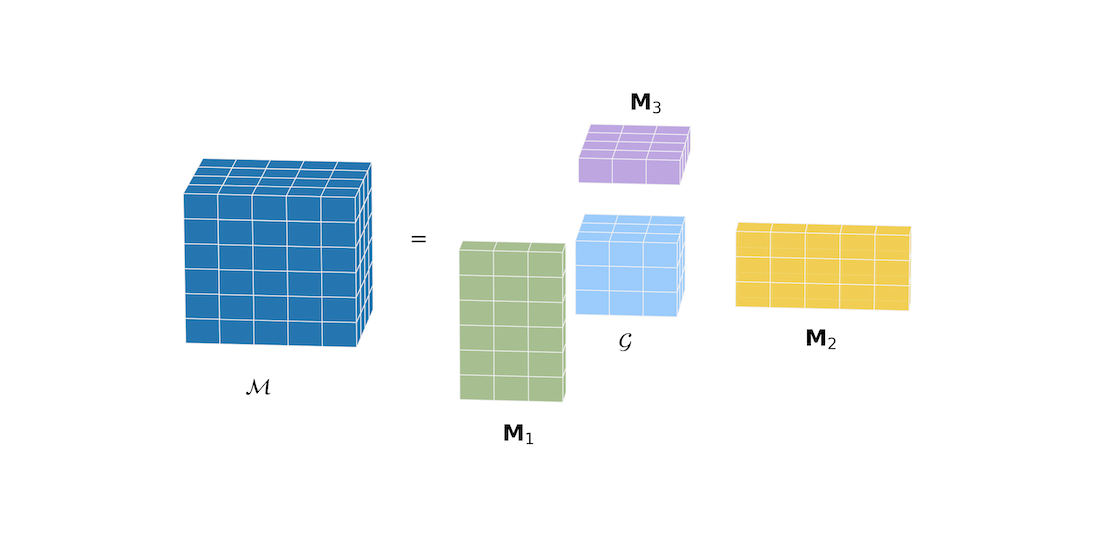}
\caption{Visualization of Three-Dimensional Tucker Decomposition}
\label{Tucker_fig}

\end{figure}

Figure \ref{Tucker_fig} illustrates the Tucker decomposition of $\mM$ when $I_1=6, I_2=5, I_3=4$, and  $R_1= R_2= R_3=3$.

Consequently, the CP and Tucker decompositions offer a much more parsimonious parameterization of $\mB$. In particular, the CP decomposition requires only
$\sum_{i=1}^{6} I_iR$
free parameters, where $I_4=I_1$, $I_5=I_2$, $I_6=I_3$, and $R$ denotes the common CP rank. Under the Tucker decomposition, the number of free parameters becomes
$
R_p + \sum_{i=1}^{6} I_i R_i,
$ where $R_p = \prod_{i=1}^{6} R_i$,
and the associated computational complexity is of order $\mathcal{O}(\max(\{I_i\}_{i=1}^3, R_p)^3)$. In a disaggregated bilateral trade application with $I_1 = I_2 = 18$ and $I_3 = 15$, both the CP and Tucker decompositions reduce the number of unknown parameters to only a few hundred, yielding substantial computational gains.

The main advantages of the Tucker decomposition over the CP decomposition lie in its flexibility in allowing different ranks across modes and in the ability of the core tensor to capture more complex cross-mode interactions.
 Moreover, the CP decomposition can be viewed as a special case of the Tucker decomposition in which the core tensor is superdiagonal and the ranks are identical across all modes.

Specifically, the Tucker decomposition of the sixth-order tensor $\mB$ can be written as
\begin{equation}\label{TKdecom}
    \mB \approx
\llbracket \mG \ ;\ \bB_1, \cdots, \bB_6 \rrbracket=
\sum_{r_1=1}^{R_1} \cdots \sum_{r_6=1}^{R_6}
g_{r_1,\cdots,r_6}
\bfb_{r_1}^{(1)} \circ \cdots \circ \bfb_{r_6}^{(6)}
\end{equation}
Equivalently, the $(i_1,\cdots,i_6)^{th}$ element of $\mB$ is given by
\begin{equation}\label{TKele}
\mB_{i_1,\cdots,i_6}
=
\sum_{r_1=1}^{R_1} \cdots \sum_{r_6=1}^{R_6}
g_{r_1,\cdots,r_6}
\bfb_{i_1,r_1}^{(1)} \times \cdots \times \bfb_{i_6, r_6}^{(6)}
\end{equation} where $\bfb_{i_1,r_i}^{(1)}$ denotes the $i_1^{th}$ element of the vector $\bfb_{r_1}^{(1)}$, and similarly for $\bfb_{i_2,r_2}^{(2)}, \cdots, \bfb_{i_6,r_6}^{(6)}$. 
The marginal vectors satisfy
$
\bfb_{r_1}^{(1)}, \bfb_{r_4}^{(4)} \in \mathbb{R}^{I_1}, \quad
\bfb_{r_2}^{(2)}, \bfb_{r_5}^{(5)} \in \mathbb{R}^{I_2}, \quad
\bfb_{r_3}^{(3)}, \bfb_{r_6}^{(6)} \in \mathbb{R}^{I_3},
$
with $r_i = 1, \dots, R_i$. The low-rank core tensor
$
\mG \in \mathbb{R}^{R_1 \times \cdots \times R_6},
$
and each loading matrix is defined as
$
\bB_i = \left[ \bfb_1^{(i)}, \cdots, \bfb_{R_i}^{(i)} \right].$
Therefore, under the Tucker decomposition, $\mB$ is approximated by the weighted sum of
$
R_p
$
rank-one tensors. 

\subsection{Identification Issue}\label{identification}
The Tucker decomposition in equation \eqref{TKdecom} suffers from an identification problem. The lack of uniqueness arises from the following factors:

1. Scale Indeterminacy: If there exists a set of scalars \(\{\lambda, \lambda_{r_1}^{(1)}, \dots, \lambda_{r_6}^{(6)}\}\), where \(r_i\) ranges from \(1\) to \(R_i\), such that \(\lambda \prod_{i=1}^6 \lambda_{r_i}^{(i)} = 1\) for any combination of \(\{r_i\}_{i=1}^6\), then replacing \(\mG\) with \(\lambda \mG\) and \(\bfb_{r_i}^{(i)}\) with \(\lambda_{r_i}^{(i)} \bfb_{r_i}^{(i)}\) results in the same decomposition.

2. Rotational Indeterminacy: For any rotation of a loading matrix $\bB_i$ ( $\bB_i$ multiplied by an $R_i \times R_i$ rotation matrix on right side), if we rotate $\mG$ inversely along mode-$i$, the decomposition remains unchanged.

We illustrate the second issue with a simple example in matrix space. Consider the previous example of an \( m \times n \) matrix \(\bM\) (a second-order tensor) of rank \( R \). In this case, the SVD would be one solution to the Tucker decomposition, as mentioned earlier, 
$
\bM = \bU \bD \bV' $,
where \(\bU\) and \(\bV\) are 
loading matrices for the first and second dimensions, respectively, and \(\bD\) is a special $R \times R$ diagonal core matrix for \(\bM\). Given two non-singular matrices \(\bP_1\) and \(\bP_2\), we can create a new pair of loading matrices:  
$\tilde{\bU} = \bU \bP_1^{-1}, \quad \tilde{\bV} = \bV \bP_2^{-1} $. 
Next, we apply an inverse rotation to the core matrix \(\bS\) along each dimension: $\tilde{\bS} = \bP_1 \bS \bP_2'.$
This yields an alternative Tucker decomposition of $\bM$. Specifically,
$\tilde{\bU} \tilde{\bS} \tilde{\bV}' = (\bU \bP_1^{-1})(\bP_1 \bS \bP_2')(\bV \bP_2^{-1})' = \bU \bS \bV' = \bM$.
Hence, the decomposition is not unique, reflecting the presence of rotation indeterminacy.

Owing to scale and rotation indeterminacy, it is desirable to select a decomposition satisfying certain normalization properties. One convenient choice is to apply the higher-order singular value decomposition (HOSVD) to the estimated, identified coefficient tensor $\mB$, as in \cite{de2000multilinear,kolda2009tensor,wang2024high}.
Through HOSVD, the new loading matrix \(\bar{\mathbf{B}}_i\)  that contains the leading \(R_i\)  left singular vectors of \(\mathbf{B}_{(i)}\)  is column orthonormal, that is, $\bar{\bB}_i'\bar{\bB}_i=\bI_{R_i}$. Furthermore, the resulting core tensor $\bar{\mG}$ is all-orthogonal, meaning \(\bar{\mathbf{G}}_{(i)} \bar{\mathbf{G}}_{(i)}^\top\) is a diagonal matrix for $i$=$1, \dots, 6$. 
 The all-orthogonality property serves an analogous role to the diagonality property in matrix SVD. In a sense, the HOSVD estimates can be viewed as the best rank-$R_i$ approximation to $\mB$ in terms of the Frobenius norm over its $i^{\text{th}}$ two-dimensional projection.

Although HOSVD estimates are still not fully identified due to potential sign indeterminacy, they have greatly mitigated the identification issue and are sufficient for our interpretation in several aspects. We can also use the projection matrices of these loading matrices, $\bar{\bB}_i\bar{\bB}_i'$, to resolve the sign indeterminacy, as demonstrated in \cite{wang2024high}.

\subsection{Global-local Shrinkage Priors}


Although the Tucker decomposition substantially reduces the number of free parameters, the resulting model may still be high-dimensional as the tensor dimensions increase. To further regularize the parameter space, we introduce shrinkage priors, which shrink negligible coefficients toward zero while preserving large coefficients, thereby improving estimation and inference efficiency. Under the low-rank Tucker decomposition, specifying a prior on \(\mB\) is equivalent to imposing priors on the associated margins or the core tensor.
A widely used class for high-dimensional problems is global-local shrinkage \citep{polson2010shrink,polson2012local}. 

Specifically, we adapt the multiway stick-breaking (SB)  shrinkage prior of \citet{guhaniyogi2020bayesian} from the CP decomposition to the Tucker framework, allowing the shrinkage to act on the margins.  In the prior specification, we assume that the ${(j,r)}^{th}$ element of the loading matrix for mode $i$, denoted by $b_{jr}^{(i)}$, follows a hierarchical distribution given by a zero-centered normal mixture:
$$ (b_{jr}^{(i)}\mid \tau_i,\phi_r^{i}) \sim \mathcal{N}\left(0, \tau_i\phi_r^{i}\right) $$  
The global parameter $\tau_i$ governs the overall degree of shrinkage in the loading matrix $\bB_i$, while the rank-specific parameters $\{\phi_r^{i}\}_{r=1}^{R_i}$ control shrinkage across ranks within mode $i$. 
Specifically, $\{\phi_r^{i}\}_{r=1}^{R_i}$ follow a stick-breaking distribution, where the sum of \(\phi_r^{i}\)  across \(r\) equals 1. 
\[
\begin{aligned}
\tau_i &\sim \operatorname{Gamma}(\alpha_{\tau_i}, \beta_{\tau_i}), \
\phi_r^{i} = \eta_r^{i} \prod_{l=1}^{r-1} \left(1 - \eta_l^i\right), \text{for } r \le R_i - 1, \
\phi_{R_i}^{i} &= \prod_{l=1}^{R_i-1} \left(1 - \eta_l^{i}\right), \\
\end{aligned}
\]
$\text{where } \eta_r^{i} \stackrel{\mathrm{iid}}{\sim} \operatorname{Beta}(1, \alpha_i)$. Conditional on $\phi_r^{i}$, $b_{jr}^{(i)}$ follows a normal–gamma distribution, which implies that its marginal prior has heavier tails than the Gaussian distribution. The increased mass concentrated at zero enhances shrinkage toward sparsity.  For the finite stick-breaking construction of the rank parameters, 
smaller values of $\alpha_i$ induce stronger shrinkage in $\phi_r^{i}$, driving them toward zero. The stick-breaking prior imposes shrinkage across ranks, promoting a low-rank and parsimonious representation. 
Hence, we do not perform explicit rank selection. Moreover, since the prior specification varies across modes $i$, it accommodates heterogeneous ranks across dimensions.  The posterior distributions of these shrinkage parameters are provided in Appendix B.

\section{Bayesian Estimation}
In this section, we develop a Bayesian estimation procedure to efficiently sample the parameters of TAR models via Markov Chain Monte Carlo (MCMC) approach under the Tucker decomposition. We first discuss the sampling of the six loading matrices, then proceed to posterior simulation of the core tensor, and finally address the covariance matrices.

\subsection{Sampling Response Loading Matrices}

In this subsection, we develop an efficient algorithm to estimate the first three response loading matrices of the TAR(1) model. We focus on the sampling procedure for $\bB_1$, while additional details of $\bB_2$ and $\bB_3$ are provided in Appendix B. We first introduce a useful proposition.



\begin{prop}\label{B123}
   Let $\bX_{it} = \by_{t-1} \otimes \bI_{k_i}$,  $k_i = \displaystyle \prod_{\substack{j=1 \\ j\ne i}}^{3} I_j$. Then the mode-\(i\) matricization of \(\mY_t\), denoted by $\bY_{t,(i)}$, can be expressed as follows:
\begin{equation}\label{TARtoMAR}
        \bY_{t,(i)}= \bA_{t,(i)}+\bB_i \bG_{(i)} \bB'_{-i}\bX_{it}+\bE_{t,(i)}, \quad  i=1,2,3
\end{equation}
\end{prop} 
where \(\bA_{t,(i)}\), \(\bE_{t,(i)}\) denote the mode-$i$ matricizations of \(\mA_t\) and \(\mE_t\), respectively, and 
$\bB_{-i}
=
\displaystyle \bigotimes_{\substack{j=6 \\ j\ne i}}^{1} \bB_j$.
 The proof of this proposition is given in Appendix A. 

This proposition allows to transform the tensor autoregressions into multiple bilinear matrix  regressions.   
The likelihood function implied by equations \eqref{eq:E_mat} and  \eqref{TARtoMAR}  for $i=1$ can be expressed as:  
 \begin{equation}\label{eq:likelihood}
 \resizebox{1\hsize}{!}{$
\begin{aligned}
P(\mY &\mid \bB_1,\bB_{-1}, \mG, \tilde{\vSigma}_1, \vomega) =
(2 \pi)^{-\frac{T I}{2}}\mid\tilde{\vSigma}_1\mid^{-\frac{T}{2}}\\
&\times \prod_{t=1}^T \omega_t^{-\frac{I}{2}} \exp\left[-\frac{1}{2\omega_t} \tvec(\bY_{t,(1)} -\bA_{t,(1)}-\bB_1\bG_{(1)}\bB'_{-1}\bX_{1t})'\tilde{\vSigma}_1^{-1}\tvec(\bY_{t,(1)}-\bA_{t,(1)}-\mathbf{B_{1}} \bG_{(1)}\mathbf{B^\prime_{-1}}  \bX_{1t}) \right]
\end{aligned}$}
\end{equation}

Let $\bfb_i=\operatorname{vec}(\bB_i)$, $i=1,2,3$, and assume the Gaussian prior
$\bfb_i \sim \mathcal{N}\!\left(\mathbf{b}_{i0},\,\mathbf{V}_{\mathbf{b}_i}\right)$, where
$\bfb_{i0}\in\mathbb{R}^{I_iR_i}$ is a zero vector, and $\bV_{\bfb_i}$ is an $I_iR_i\times I_iR_i$ diagonal matrix of shrinkage prior variances. It follows that 
 \begin{equation}\label{eq:B1}
 \resizebox{1\hsize}{!}{$
\begin{aligned}
P(\bfb_1 \mid \mY,\{\bB_i\}_{i\ne 1}, \mG, \tilde{\vSigma}_1, \vomega) &\propto   \exp\left[-\frac{1}{2} (\bfb_1-\mathbf{b}_{i0})'\bV_{\bfb_1}^{-1}(\bfb_1-\mathbf{b}_{i0})\right]\\
&\times \prod_{t=1}^T \exp\left[-\frac{1}{2\omega_t}\tvec(\bY_{t,(1)} -\bA_{t,(1)}-\bB_1\bG_{(1)}\bB'_{-1}\bX_{1t})'\tilde{\vSigma}_1^{-1}\tvec(\bY_{t,(1)}-\bA_{t,(1)}-\mathbf{B_{1}} \bG_{(1)}\mathbf{B^\prime_{-1}}  \bX_{1t})\right]
 \\&\propto   \exp\left[-\frac{1}{2} \left(\bfb_1'\bV_{\bfb_1}^{-1} \bfb_1 +\sum_{t=1}^T\omega^{-1}_t \bfb_1'(\bG_{(1)}\mathbf{B^\prime_{-1}}  \bX_{1t}\otimes \bI_{I_1})\tilde{\vSigma}_1^{-1}(\bX'_{1t}\mathbf{B_{-1}}  \bG'_{(1)}\otimes \bI_{I_1}) \bfb_1\right)\right]\\ &\times \exp\left[-\sum_{t=1}^T\bfb_1'(\bV_{\bfb_1}^{-1}\mathbf{b}_{i0}+\omega^{-1}_t(\bG_{(1)}\mathbf{B^\prime_{-1}}  \bX_{1t}\otimes \bI_{I_1})\tilde{\vSigma}_1^{-1}\tvec(\bY_{t,(1)}-\bA_{t,(1)})\right]
  \\&\propto   \exp\left[-\frac{1}{2} \left(\bfb_1'\bV_{\bfb_1}^{-1} \bfb_1 +\sum_{t=1}^T \bfb_1'(\omega^{-1}_t\bG_{(1)}\mathbf{B^\prime_{-1}}  \bX_{1t}(\vSigma_3^{-1}\otimes \vSigma_2^{-1})\bX'_{1t}\mathbf{B_{-1}}  \bG'_{(1)}\otimes \vSigma^{-1}_{1}) \bfb_1\right)\right]
  \\ &\times \exp\left[-\sum_{t=1}^T\bfb_1'(\bV_{\bfb_1}^{-1}\mathbf{b}_{i0}+\omega^{-1}_t(\bG_{(1)}\mathbf{B^\prime_{-1}}  \bX_{1t}(\vSigma_3^{-1}\otimes \vSigma_2^{-1})\otimes\vSigma_1^{-1})\tvec(\bY_{t,(1)}-\bA_{t,(1)}))\right]
   \\&\propto   \exp\left[-\frac{1}{2} \left(\bfb_1'\bK_{\bfb_1} \bfb_1-2\sum_{t=1}^T\bfb_1'(\bV_{\bfb_1}^{-1}\mathbf{b}_{i0}+\omega^{-1}_t\tvec(\vSigma_1^{-1}(\bY_{t,(1)}-\bA_{t,(1)})(\vSigma_3^{-1}\otimes \vSigma_2^{-1})\bX_{1t}'\bB_{-1}\bG_{(1)}')) \right)\right]
    \\&\propto   \exp\left[-\frac{1}{2} \left(\bfb_1'\bK_{\bfb_1} \bfb_1-2\sum_{t=1}^T\bfb_1'\bK_{\bfb_1}\hat{\bfb}_1\right)\right]
\end{aligned}$}
\end{equation}
where 
$$
\begin{aligned}\bK_{\bfb_1}=&\bV_{\bfb_1}^{-1}+\sum_{t=1}^T \omega_t^{-1}\left(\bG_{(1)}\bB_{-1}^{\prime} \bX_{1t}(\vSigma_3^{-1}\otimes \vSigma_2^{-1}) \bX_{1t}^{\prime}\bB_{-1}\bG_{(1)}'\right)\otimes \vSigma_1^{-1} \\ \hat{\bfb}_1=&\bK_{\bfb_1}^{-1}\left(\bV_{\bfb_1}^{-1} \bfb_{10}+\sum_{t=1}^T \omega_t^{-1}\tvec\left(  \vSigma_1^{-1}(\bY_{t, (1)}-\bA_{t,(1)})(\vSigma_3^{-1}\otimes \vSigma_2^{-1})\bX_{1t}'\bB_{-1}\bG'_{(1)} \right)\right)\end{aligned}$$

Equivalently, \begin{equation}\label{B_1p}(\bfb_1\mid\mY,\{\bB_i\}_{i\ne 1},\mG, \tilde{\vSigma}_1, \vomega)  \sim \distn{N} (\widehat{\mathbf{b}}_1,\mathbf{K}_{\mathbf{b}_1}^{-1}) \end{equation}



Similarly, as shown in Appendix B, \((\bfb_2 \mid \mY, \{\bB_i\}_{i\ne 2}, \mG, \tilde{\vSigma}_2, \vomega)\) and \((\bfb_3 \mid \mY, \{\bB_i\}_{i\ne 3}, \mG, \tilde{\vSigma}_3, \vomega)\) also follow Gaussian distributions.

\subsection{Sampling Predictor Loading Matrices}
 We next derive the posterior distributions of the predictor loading matrices, $\bB_4, \bB_5$, and $\bB_6$. Compared with the response loading matrices, their derivations involve additional technical challenges.
 Our discussion focuses on \(\bB_4\) in the main text, while the Gibbs sampler of \(\bB_5\) and \(\bB_6\) is discussed in Appendix B. The key step is to isolate \(\bB_4\) from other Tucker components, which can be achieved via the following proposition.


\begin{prop}\label{MVAR}
The proposed TAR(1) model can be expressed as a special case of a VAR(1) model, with the coefficient matrix parameterized by the Tucker components:
\begin{equation}\label{eq:MVAR}
  \by_t=\ba_t+\tilde{\bB}_m\tilde{\bG}\tilde{\bB}'\by_{t-1}+\be_t, \quad \be_t \sim \mathcal{N}(\mathbf{0}_I, \omega_t\vSigma)
\end{equation}
\end{prop}
where  $\tilde{\bB}_m=\bB_3\otimes\bB_2\otimes\bB_1$, 
 $\tilde{\bB}=\bB_6\otimes\bB_5\otimes\bB_4$. $\tilde{\bG}$ is the $\tilde{R}_m\times\tilde{R}$ matricization of the core tensor $\mG$, satisfying $\tvec(\tilde{\bG})=\tvec(\mG)$, where $\tilde{R}_m=R_1R_2R_3$ and $\tilde{R}=R_4R_5R_6$.
The proof of this proposition is provided in Appendix A.

Compared to equation \eqref{eq:VAR1}, Proposition \ref{MVAR} shows that the TAR(1) model under Tucker decomposition can be represented as a special VAR(1) model, where the $I \times I$ coefficient matrix is given by $\hat{\bB}_1 = \tilde{\bB}_m \tilde{\bG} \tilde{\bB}'$. The lagged effect of $\mY_{t-1,i_4,i_5,i_6}$ on $\mY_{t,i_1,i_2,i_3}$ is captured by the $(k_1,k_2)^{th}$ element of $\hat{\bB}_1$, where $k_1 = i_1 + (i_2 - 1) I_1 + (i_3 - 1) I_1 I_2$ and $k_2 = i_4 + (i_5 - 1) I_1 + (i_6 - 1) I_1 I_2$.

Proposition $ \ref{MVAR} $ reduces the estimation problem from a three-dimensional setting to a classical vector regression and facilitates the separation of $ \bB_4 $ from other loading matrices. To this end, since we have 
\begin{equation*}
\tilde{\bB}'\by_{t-1}=\tilde{\bB}'\tvec(\bY_{t-1,(1)})=\tvec\left(\bB'_4\bY_{t-1,(1)}(\bB_6\otimes\bB_5)\right)=\left((\bB'_6\otimes\bB'_5)\bY_{t-1,(1)}'\otimes\bI_{R_4}\right)\tvec(\bB'_4)
\end{equation*} 
Let $\bfb_4=\operatorname{vec}(\bB'_4)$. Equation \eqref{eq:MVAR} can therefore be rewritten as follows:
\begin{equation}\label{eq:B4}
\by_t=\ba_t+\tilde{\bB}_m\tilde{\bG}\left((\bB'_6\otimes\bB'_5)\bY_{t-1,(1)}'\otimes\bI_{R_4}\right)\bfb_4+\be_t
\end{equation}
Suppose the prior on $\bfb_4$ is also given by a Gaussian distribution, $\bfb_4  \sim \distn{N}(\bfb_{40},\bV_{\bfb_4})$,
where $\bfb_{40}$ is a zero vector of dimension $I_1R_4$, and $\bV_{\bfb_4}$ is an $I_1R_4\times I_1R_4$ diagonal matrix with shrinkage prior variances. Combined with the likelihood implied by equation $ \eqref{eq:B4} $, the posterior distribution of $ \bfb_4 $ can be expressed as follows:
$$(\bfb_4|\mY, \{\bB_i\}_{i\ne 4},\bG, \vSigma, \vomega) \sim \distn{N}(\hat{\bfb}_4,\bK^{-1}_{\bfb_4})$$
where $$
\begin{aligned}
\bK_{\bfb_4}=&\bV_{\bfb_4}^{-1}+\sum_{t=1}^T{\omega_t^{-1}}\left(\bY_{t-1,(1)}(\bB_6\otimes\bB_5)\otimes\bI_{R_4}\right)\tilde{\bG}'\tilde{\bB}'_m\vSigma^{-1}\tilde{\bB}_m\tilde{\bG}\left((\bB'_6\otimes\bB'_5)\bY_{t-1,(1)}'\otimes\bI_{R_4}\right)\\ \hat{\bfb}_4=&\bK_{\bfb_4}^{-1}\left(\bV^{-1}_{\bfb_4}\bfb_{40}+\sum_{t=1}^T\omega_t^{-1}\left(\bY_{t-1,(1)}(\bB_6\otimes\bB_5)\otimes\bI_{R_4}\right)\tilde{\bG}'\tilde{\bB}'_m\vSigma^{-1}(\by_t-\ba_t)\right)\end{aligned}$$
The samplers for \(\bB_5\) and \(\bB_6\) are derived similarly but involve more cumbersome algebra. Their full conditional distributions are presented in Appendix B.



\subsection{Sampling the Core Tensor $\mG$}
With respect to $\mG$, consider the stacked representation of equation $ \eqref{eq:MVAR} $ obtained by defining $\bY = (\by_1, \ldots, \by_T)$, $\bA = (\ba_1, \ldots, \ba_T)$, $\bX = (\by_0, \ldots, \by_{T-1})$, and $\bE = (\be_1, \ldots, \be_T)$. It follows that
$$
\bY = \bA + \tilde{\bB}_m \tilde{\bG} \tilde{\bB}' \bX + \bE
$$
Vectorizing both sides gives
\begin{equation}
\begin{aligned}
\by=\ba+(\bX'\tilde{\bB}\otimes\tilde{\bB}_m)\tvec(\tilde{\bG})+\be=\ba+(\bX'\tilde{\bB}\otimes\tilde{\bB}_m)\tvec(\mG)+\be
\end{aligned}
\end{equation} 
where $\by=(\by_1',\cdots,\by_T')'$, $\ba=(\ba_1',\cdots,\ba_T')'$, $\be=(\be_1',\cdots,\be_T')'$. Suppose that the prior on $\bg=\tvec(\mG)$ is given by $\bg \sim \distn{N}(\bg_0,\bV_{\bg})$,
where $\bg_0\in\mathbb{R}^{R}$ is a zero vector, and $\bV_{\bg}$ is an $R\times R$ diagonal matrix. Under standard linear regression theory, the posterior distribution of $ \bg $ has the following form:
$$\left(\bg|\mY, \{\bB_i\}_{i=1}^6,\vSigma, \vomega\right) \sim \distn{N}(\hat{\bg},\bK^{-1}_{\bg})$$
where $$\bK_{\bg}=\bV_{\bg}^{-1}+\left(\tilde{\bB}'\bX\vOmega^{-1}\bX'\tilde{\bB}\right)\otimes\left(\tilde{\bB}_m'\vSigma^{-1}\tilde{\bB}_m\right)$$
$$\hat{\bg}
=\bK_{\bg}^{-1}\left(\bV^{-1}_{\bg}\bg_0+\tvec(\tilde{\bB}_m'\vSigma^{-1}(\bY-\bA)\vOmega^{-1}\bX'\tilde{\bB})\right)$$

\subsection{Sampling the Covariance Matrices}
For the covariance matrices $\vSigma_i$, $i=1,2,3$, assume the inverse Wishart prior
$
\vSigma_i \sim \mathcal{IW}(\nu_i,\mathbf{S}_{i0}).
$ It then follows that the posterior distribution of $\vSigma_i$ remains inverse Wishart, given by:
\[
\vSigma_i \sim \mathcal{IW} \left(\nu_i + TI_{i_1}I_{i_2}, \widehat{\mathbf{S}}_i \right).
\]
where $\hat{\bS}_i=\bS_{i0}+\sum_{t=1}^T \omega_t^{-1} \bE_{t, (i)}(\vSigma_{i_1}^{-1}\otimes \vSigma_{i_2}^{-1}) \bE_{t, (i)}^{\prime}$, for $ i_1, i_2 \in \{1,2,3\} \setminus {i} $ with $ i_1 > i_2 $.

This section presents the Bayesian estimation procedure for the Tucker components of the coefficient tensor and the mode-specific covariance matrices. The following sections evaluate the proposed method through Monte Carlo simulations and an empirical application to global trade data, illustrating its ability to identify influential countries and commodity groups and to uncover the latent factors driving global trade dynamics.

\section{Simulation Results}\label{simulation}

In this section, we examine the finite-sample performance of the proposed model by comparing four estimators: (1) a Bayesian VAR(1) estimator with a Minnesota prior (BVAR-Minn), (2) a CP decomposition-based estimator (BTAR-CP), (3) a Tucker decomposition-based estimator (BTAR-TK), and (4) a Tucker decomposition-based estimator incorporating the multiway stick-breaking shrinkage prior (BTAR-TK-Sh). Note that the CP decomposition can be viewed as a special case of the Tucker decomposition in which the core tensor is superdiagonal and the ranks are identical across all modes. Accordingly, for the CP specification, we restrict the core tensor to remain superdiagonal throughout the MCMC iterations and impose a common rank across all modes, such that $R_i = R$ for $i = 1, \ldots, 6$. We denote this specification by CP($R$).

Specifically, we consider two simulation experiments. In the first experiment, the coefficient tensor in the data-generating process (DGP) is characterized by an underlying low-rank structure. In the second experiment, the coefficient tensor is generated from a Gaussian distribution without imposing any low-rank structure. For each experimental setting, the results are based on 20 simulation replications.

In the first experiment, we assume that $\mB$ follows a low-rank Tucker decomposition as specified in equation $\eqref{TKdecom}$, and use this structure to generate the data using the equation \eqref{BTAR1}.
Each element of the core tensor $\mG$ follows a uniform distribution $\distn{U}(0, 1)\) and the marginal vectors \(\bfb^{(i)}_r \sim \distn{N}(0.3, 0.5^2 \bI_{I_i})\). Additionally, $\mB$ is normalized so that its Frobenius norm\footnote{The Frobenius norm of a tensor $\mB$ is defined as $ \| \mB \| =\sqrt{\langle \mB, \mB \rangle}.$} is equal to 5. For simplicity, all entries of $\mA_0$ and $\mA_1$ are set to zero. The error terms \(\boldsymbol{\epsilon}_t\) are generated in accordance with equation \eqref{error} with \(\omega_t = 1\) for all \(t\).  The error covariance matrices $\vSigma_1 $,  $\vSigma_2 $, and $\vSigma_3 $ are drawn from $ \distn{IW}(I_1 + 10; \bI_{I_1}) $, $\distn{IW}(I_2 + 10; \bI_{I_2})$, and  $\distn{IW}(I_3 + 10; \bI_{I_3})$, respectively, and then each scaled by the sum of their diagonal entries.  The DGP cases differ in dimensions—either $I_1 = I_2 = I_3 = 5$ or $I_1 = I_2 = 10, I_3 = 2$—with average estimation errors reported for $T = 100, 200, 300$ or $T = 200, 300, 400$, respectively.  Additionally, two Tucker rank settings are used: (1) $R_i = 2$ for all $i$; (2) $(R_1,\ldots,R_6) = (1,1,1,2,2,2)$. In Case 2, since CP models cannot accommodate unequal ranks across modes, we instead use an alternative CP model with $R_i = 2$ for all $i$.  We consider both dense and sparse data-generating processes, where in the sparse specification, the last columns of $\bB_1$ and $\bB_4$, together with half of the elements of $\mB$, are fixed at zero.

To implement the Bayesian VAR(1) estimator, we first vectorize the tensor $\mY_t$ by stacking all of its elements into a vector $\by_t$, and then estimate a conventional VAR(1) model for $\by_t$. Unlike the CP- and Tucker-based estimators, the BVAR(1) estimator does not impose a low-rank structure on the coefficient tensor and therefore provides a more flexible representation of the dynamic relationships among the elements of $\mY_t$. This additional flexibility, however, comes at the cost of a substantially larger number of parameters, which can lead to high estimation uncertainty in finite samples, especially as the dimensions of the tensor increase. Consequently, relative to the CP- and Tucker-based estimators, the BVAR(1) estimator generally leads to lower bias but higher variance, illustrating the bias–variance tradeoff between an unrestricted VAR specification and low-rank tensor representations.

\begin{figure}[H]

\centering
\includegraphics[width=1.05\textwidth]{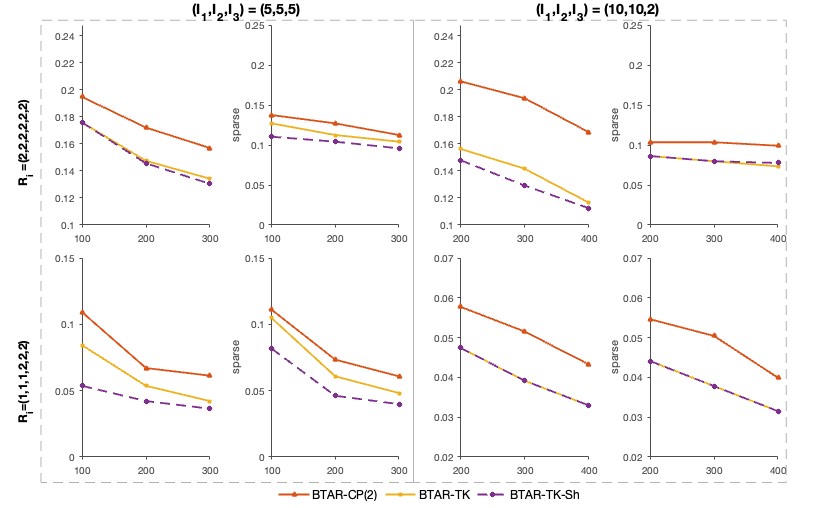} 
\caption{Relative RMSEs under Low-Rank Data-Generating Processes}
\label{simu_fig}

\end{figure}

Figure \ref{simu_fig} presents the simulation results from the first experiment, showing the relative root mean squared errors (rRMSEs) of the low-rank models, which are computed by dividing each model's RMSE by the minimum RMSE attained by the benchmark Bayesian VAR(1) model across all sample sizes considered. As shown in the figure, the rRMSEs remain below one across all specifications, demonstrating that the proposed low-rank models consistently achieve greater estimation accuracy than the Bayesian VAR(1) benchmark. Moreover, the rRMSEs decline as the sample size increases, indicating improved estimation accuracy with larger sample sizes. Across different model dimensions and rank specifications, the Tucker-based BTAR models consistently outperform their CP-based counterparts. This advantage remains evident under the rank specification $(R_1,\ldots,R_6)=(1,1,1,2,2,2)$, even when CP models with $R_i=2$ are chosen as a comparison. Overall, these findings highlight the greater flexibility of the Tucker decomposition relative to the CP decomposition. In addition, the proposed shrinkage prior generally yields a favorable bias--variance tradeoff, leading to improved estimation accuracy under both sparse and dense data-generating processes.



In the second experiment, we consider a more general setting in which each element of $\mB$ is generated from an independent Gaussian distribution without imposing a low-rank structure. We then generate data according to the TAR(1) model in equation \eqref{BTAR1}, or, equivalently, its vectorized VAR(1) representation in equation \eqref{eq:VAR1}.
Specifically, consider the coefficient matrix $\hat{\bB}$ in equation \eqref{eq:VAR1}, which corresponds to the matrix representation of the coefficient tensor $\mB$. The off-diagonal elements of $\hat{\bB}$ are independently drawn from a normal distribution, $\mathcal{N}(0,0.2^2)$, while the diagonal elements are drawn from a uniform distribution, $U(0.1,0.3)$. The resulting matrix $\hat{\bB}$ is then normalized to have unit Frobenius norm. The intercept terms $\ba_t$ and error terms $\be_t$ are generated in the same manner as in the first experiment. Using the resulting VAR(1) representation, we generate $\by_t$ and reshape it into a third-order tensor $\mY_t$ such that $\tvec(\mY_t)=\by_t$. The resulting tensor observations $\mY_t$ are then used to estimate the coefficient tensor $\mB$ under the low-rank tensor specifications. 

We consider both matrix autoregressions (MARs), corresponding to two-dimensional tensor autoregressions, and third-order tensor autoregressions (TARs), with a sample size of $T=500$. For each model specification, we consider two dimensional settings. For MARs, we set $(I_1,I_2)=(5,5)$ and $(15,5)$, while for TARs, we consider $(I_1,I_2,I_3)=(3,3,4)$ and $(3,4,5)$. As in the previous experiment, we examine both dense and sparse settings. For both the CP and Tucker decompositions, we consider ranks $R_i=1,2,3,4$ for MARs and $R_i=1,2,3$ for TARs, with the same rank imposed across all modes. The reported estimation errors are averaged across simulation replications.

\begin{figure}[H]

\centering
\includegraphics[width=1.05\textwidth]{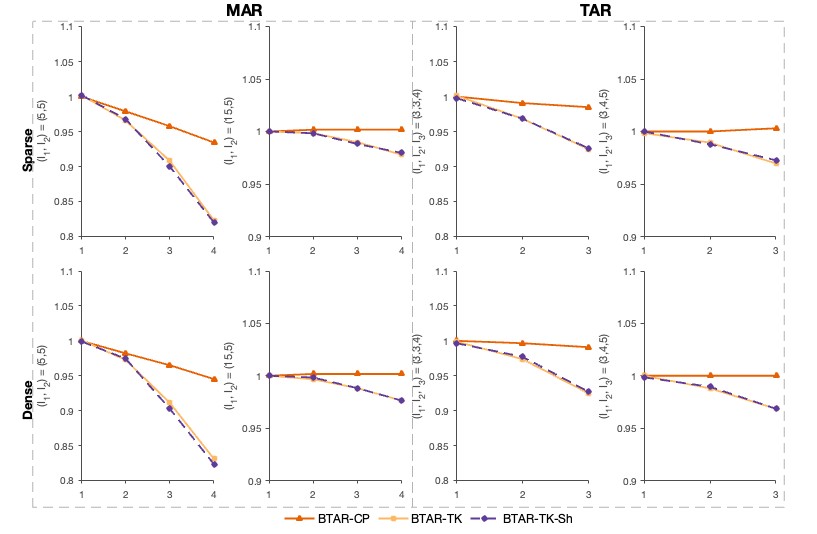} 
\caption{Relative RMSEs under Non-Low-Rank Data-Generating Processes}
\label{simu_var}

\end{figure}

Figure \ref{simu_var} presents the relative root mean squared errors (rRMSEs) across different rank specifications, as shown on the x-axis, with the RMSE of the CP model with $R_i=1$ used as the normalization benchmark. When $R_i=1$, the CP and Tucker decompositions are theoretically equivalent because the Tucker core tensor reduces to a scalar, a result that is also reflected in the figure. As the rank increases, the rRMSEs of both the MAR and TAR models generally decline, suggesting that higher-rank specifications are able to capture a larger share of the underlying dynamic structure. Across all rank specifications, the Tucker-based models consistently outperform the corresponding CP models, with the performance gap widening as the rank increases. This pattern highlights the greater flexibility of the Tucker decomposition in approximating coefficient structures that do not have an underlying low-rank structure. Moreover, the Tucker models with shrinkage priors achieve estimation accuracy that is generally comparable to, and often better than, their non-shrinkage counterparts, indicating a favorable bias--variance tradeoff as well. 

It is worth emphasizing that the data-generating process considered in the second experiment is deliberately challenging for the proposed low-rank models, as the coefficients are generated independently without imposing any underlying low-rank structure. Despite this unfavorable setting, estimation errors decline steadily as the rank increases, demonstrating that higher-rank specifications allow the proposed framework to approximate increasingly complex coefficient structures more accurately.

\section{Empirical Application}

In this section, we analyze monthly bilateral trade flow data from the UN Comtrade Database. The dataset covers 18 countries across 15 commodity categories over the period January 2010 to December 2023. The sample includes one Asian economy (Japan), two North American economies (the United States and Canada), one South American economy (Brazil), and fourteen European economies (Germany, the United Kingdom, France, Italy, the Netherlands, Spain, Poland, Sweden, Belgium, Ireland, Norway, Denmark, Portugal, and Turkey). The fifteen commodity groups cover a broad range of goods and are classified according to two-digit Harmonized System (HS) codes, as detailed in Appendix C.

To mitigate the influence of incidental trades with unusually large or small values, we smooth trade values using three-month moving averages and subsequently transform them into normalized year-over-year growth rates. By doing so, we actually remove scale effects and instead focus on structural patterns to analyze systematic trade dominance.

As a result, the $18 \times 18 \times 15$ tensor $\mY_t$ comprises 4,860 time series with a sample size of $T = 156$, covering the period from January 2011 to December 2023. The element $\mY_{t,i_1,i_2,i_3}$ denotes the normalized, three-month–averaged growth rate of imports of commodity $i_3$ from exporting country $i_2$ to importing country $i_1$. The coefficient $\mB_{i_1,i_2,i_3,j_1,j_2,j_3}$ captures the effect of the one-month lagged import growth of commodity $j_3$ from country $j_2$ to country $j_1$ on the current growth of commodity $i_3$ between countries $i_1$ and $i_2$.

We employ Tucker decomposition with stick-breaking shrinkage priors for estimation. We set the lag order as $p = 1$ and the ranks to $R_i = 3$ for all modes.
 As discussed above, the prior adaptively imposes rank-specific shrinkage that favors low-rank structures; therefore, we do not impose an explicit rank-selection step.



 \subsection{Common Stochastic Volatility}
 
\begin{figure}[H]

\centering
\includegraphics[width=1\textwidth]{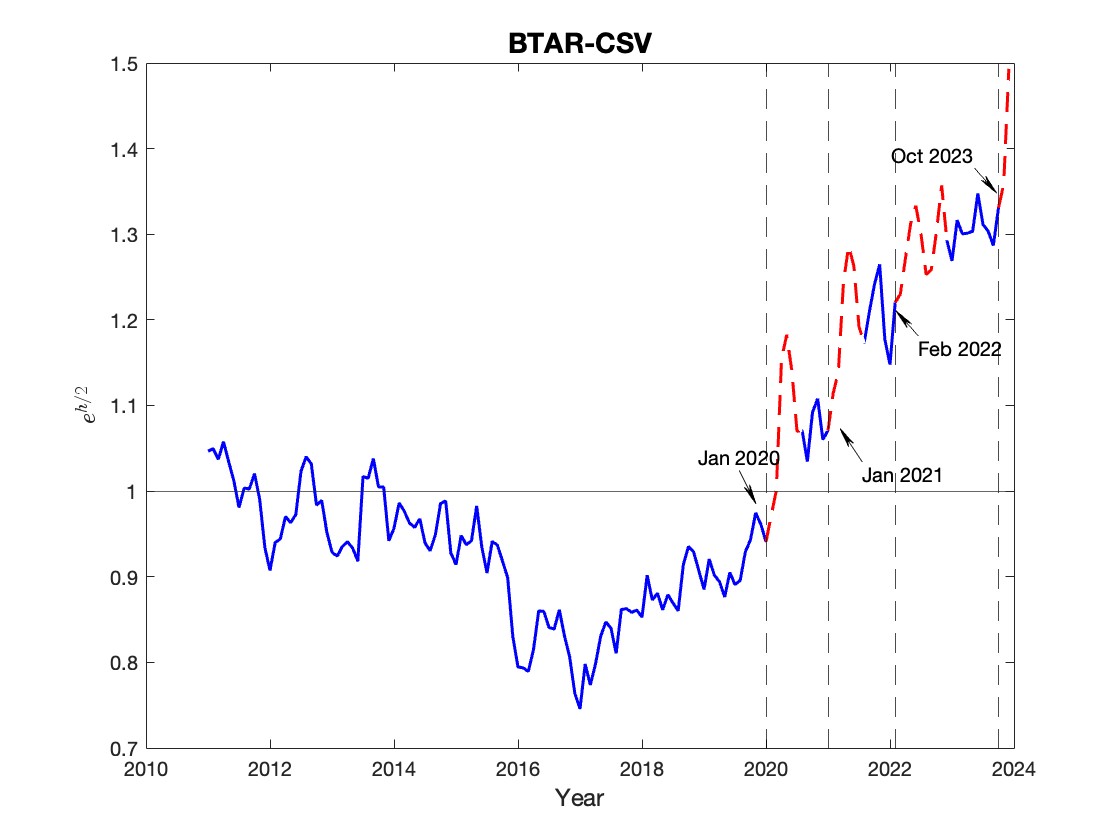} 
\caption{Posterior Means of Time-Varying Standard Derivation, $exp(h_t/2)$, from BTAR-CSV Model}
\label{omega_fig}

\end{figure}

 Figure \ref{omega_fig} presents the posterior means of the time-varying standard deviations from the BTAR-CSV model. The figure reveals several notable fluctuations, indicated by the red dotted lines, with the first spike occurring at the start of the COVID-19 pandemic in January 2020 which unleashed a global economic shock, leading to the most severe worldwide crisis in over a century,  followed by a bounce-back one year later as the world recovered from the pandemic. The third jump took place in February 2022 during the onset of the Russia-Ukraine war. The conflict, along with ensuing sanctions, severely disrupted regional exports of multiple commodities—including metals, food, oil, and gas—triggering the highest inflation in decades. The final surge reflects escalated regional tensions following the outbreak of the Israel-Hamas war in October 2023. In its aftermath, attacks in the Red Sea prompted shipping companies to avoid the Suez Canal, rerouting via the Cape of Good Hope—adding roughly 3,500 nautical miles and 14 days to voyages. This has especially disrupted trade between Europe and Asia.


\subsection{ Interpretation of $\bB_i$}

We can interpret the matrices $\bB_i$, for $i=1,\dots,6$, from the perspective of two-way factor analysis. To illustrate, consider $\bB_1$. First, revisit equation $\eqref{TARtoMAR}$ for $i=1$. Define an $R_1 \times I_2 I_3$ matrix $\bP_{1t} = \bG_{(1)} \bB'_{-1} \bX_{1t}$. Then, the mode-1 tensor slice \(\mY_{t,\cdot,i_2,i_3}\) can be expressed as:
\[
     \mY_{t,\cdot,i_2,i_3} = \mA_{t,\cdot,i_2,i_3}+ \bB_1 \bP_{1t,\cdot, (i_3-1) I_2 + i_2} + \mE_{t,\cdot,i_2,i_3}.
\]
\(\bB_1\) can be interpreted as the loading matrix in a standard factor model for all mode-1 fibers \(\mY_{t,\cdot,i_2,i_3}\), across all combinations of \((t, i_2, i_3)\). Specifically, \(\mY_{t,\cdot,i_2,i_3}\) represents the import values of commodity \(i_3\) from source country \(i_2\) to all destination countries at time \(t\), and is fitted using a $R_1 \times 1$ factor $\bP_{1t,\cdot, (i_3-1) I_2 + i_2}$. Similarly, \(\bB_2\) and \(\bB_3\) serve as loading matrices for the standard factor models of all mode-2 fibers \(\mY_{t,i_1,\cdot,i_3}\) and mode-3 fibers \(\mY_{t,i_1,i_2,\cdot}\), respectively.

An alternative factor interpretation is as follows. As discussed in Subsection~$\ref{identification}$, the HOSVD ensures that each $\bB_i$ satisfies the orthonormality condition $\bB_i' \bB_i = \bI_{R_i}$. Therefore, it follows from equation~$\eqref{eq:MVAR}$ with $\ba_t=\mathbf{0}$ that
\[
\tilde{\bB}_m' \by_t = \tilde{\bG} \tilde{\bB}' \by_{t-1} + \tilde{\bB}_m' \be_t,
\]
This formula allows to provide a dynamic factor interpretation on both sides.  In particular, the $\tilde{R}=\prod_{i=4}^6 R_i $ lagged (predictor) factors, \(\tilde{\bB}'\by_{t-1}\), drive the dynamics of the international bilateral trade flows. Meanwhile, the current state of the trade market is captured by the \(\tilde{R}_m=\prod_{i=1}^3 R_i\) response factors, \(\tilde{\bB}_m' \by_t\). $\tilde{\bG}$, the matricization of compact core tensor \(\mG\), encapsulates the predictive relationships between these lower-dimensional predictors and responses. Specifically, each loading matrix projects high-dimensional, disaggregated predictors or responses onto a specific lower-dimensional subspace, as shown in the following equations:
\begin{equation*}
    \tilde{\bB}_m' \by_t=\tvec\left(\bB_1'\bY_{t,(1)}(\bB_3\otimes\bB_2)\right)
\end{equation*}
and 
\begin{equation*}
    \tilde{\bB}' \by_{t-1}=\tvec\left(\bB_4'\bY_{t-1,(1)}(\bB_6\otimes\bB_5)\right)
\end{equation*}

In the bilateral trade flow application, $\bB_1$ (and $\bB_4$) represents the factor loadings that project the responses (or predictors) onto the first mode, corresponding to the import dimension. Similarly,  $\bB_2$ (and $\bB_5$), and $\bB_3$ (and $\bB_6$) contain the factor loadings for the export and goods dimensions of the responses (or predictors), respectively.

\begin{figure}[H]
\centering
\includegraphics[width=1\textwidth]{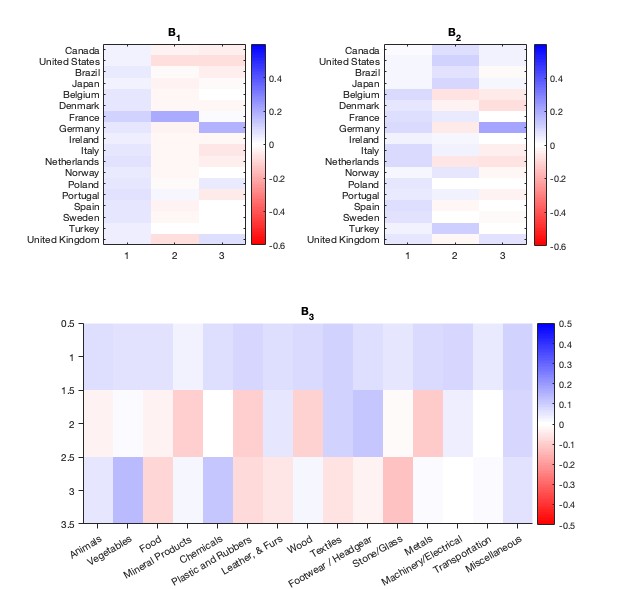} 
\caption{Heatmaps of Response Loading Matrices, $\bB_1$, $\bB_2$ and $\bB_3$}
\label{resp_fact_1}
\end{figure}

\begin{figure}[H]
\centering
\includegraphics[width=1\textwidth]{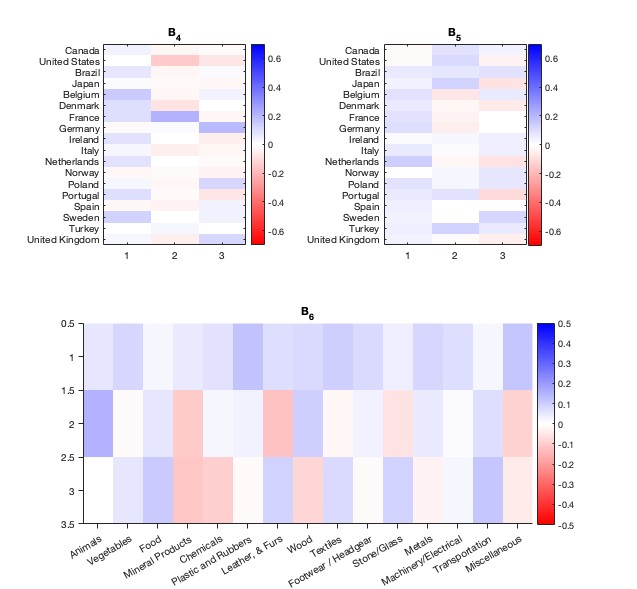} 
\caption{Heatmaps of Predictor Loading Matrices, $\bB_4$, $\bB_5$ and $\bB_6$}
\label{pred_fact_1}
\end{figure}

Figures \ref{resp_fact_1} - \ref{pred_fact_1} display heatmaps of the estimated loading matrices \footnote{Note that $\bB_i$ is subject to sign indeterminacy in each column. We choose the sign so that the element with the largest loading in each column is positive.}, \(\bB_i\), $i=1,2,4,5$, and of the projection matrices \(\bB_i \bB_i'\), $i=3,6$, from the Tucker model with hierarchical shrinkage priors on the margins. The projection matrix is identifiable, and its diagonal elements measure the total loading strength of each commodity across all factors.

Specifically, Figure~\ref{resp_fact_1} presents the estimated factor structures for the response dimensions, including the importer-country ($\bB_1$), exporter-country ($\bB_2$), and commodity ($\bB_3$) dimensions. Since Tucker factors are not uniquely identified with respect to sign, the signs of factor loadings should be interpreted relative to other elements within the same factor rather than as having absolute economic meanings. Countries or commodities with loadings of the same sign share similar positions in the corresponding latent factor space, whereas opposite signs indicate contrasting patterns along that dimension. The importer-country factor loadings in \(\bB_1\) reveal both global and country-specific patterns. The first factor has positive and non-negligible loadings across most countries, indicating a common importer factor. The second and third factors capture deviations from this common component. Specifically, the second factor is dominated by a strong positive loading for France and a negative loading for the U.S. and U.K. , distinguishing France from other two major economies in terms of their importer dynamics. In contrast, the third factor is dominated by Germany, with the U.K. also showing a positive loading, whereas the U.S., Italy and several other countries have negative loadings. These patterns suggest that the importer heterogeneity is primarily driven by contrasts among a small number of major economies rather than by broad regional differences. As for the exporter-country factor loadings in $\bB_2$ , the first factor places relatively larger weights on most European economies. In contrast, the second factor loads positively on a mix of non-European economies and other European countries, and tend to assign negative loadings to those countries that receive relatively large positive weights in the first factor, such as Belgium, Germany and Netherlands, suggesting a complementary geographic pattern across the two factors. The third factor is dominated by Germany, highlighting its significant role in the latent exporter structure.

The commodity factor loadings in $\bB_3$ reveal latent structures across commodity categories. The first commodity factor assigns positive loadings to most categories, suggesting a common factor across commodities. The second factor distinguishes consumer- and light-manufacturing goods, such as textiles, footwear and headgear, and miscellaneous products, from resource- and material-intensive commodities, including mineral products, wood, plastics and rubbers, and metals. The third factor captures an additional dimension of commodity-specific heterogeneity. Its loadings are dispersed across commodity categories, showing that this latent dimension reflects further heterogeneity in the underlying trade dynamics.

As shown in Figure \ref{pred_fact_1}, the predictor-side factor loadings characterize the latent structures embedded in lagged trade conditions. The lagged importer factors in \(\bB_4\) also reveal both common and country-specific patterns. The first factor assigns positive loadings to a broad set of countries, suggesting a global importer factor in the lagged trade structure. The second and third factors capture country-specific heterogeneity beyond the common component identified by the first factor. Whereas the second factor contrasts countries such as the U.S. and France through loadings of opposite signs, the third factor is highly concentrated on Germany, with only modest loadings for several other countries, such as the U.K., indicating a latent dimension largely driven by Germany's distinctive predictor dynamics. In terms of the lagged exporter factors in \(\bB_5\), the first factor still represents a global factor. The second factor distinguishes countries with different exporter profiles, placing heavy loadings on those non-European economies and Turkey—a transcontinental country spanning Europe and Asia—while other European economies exhibit weak or negative loadings along this dimension. The third factor captures a residual dimension of heterogeneity beyond the common factor represented by the first factor and the contrast identified by the second factor. Its loadings are distributed across countries from different regions, indicating that this factor reflects country-specific predictor dynamics rather than broad geographic patterns.

Finally, we turn to the lagged commodity factor loadings in $\bB_6$. As in $\bB_3$, the first factor still represents a common trend. The second and third factors capture additional heterogeneity by contrasting different groups of commodities. In particular, mineral products consistently have loadings opposite in sign to several other commodity groups, particularly agricultural products and transportation. By contrast, machinery and electrical products, plastics and rubbers, and footwear have relatively small loadings on these two factors, indicating that their variation is largely explained by the common component. Commodity groups such as wood and leather and furs display contrasting loading patterns across the second and third factors, suggesting that they are influenced by multiple latent dimensions rather than a single dominant factor.

\subsection{Interpretation for Predictor and Response Factor Time Series}

In this subsection, we provide a dynamic factor interpretation of both the response and predictor components. Specifically, we construct $R_1R_2R_3=27 $ response factors, $\tilde{\bB}_m^{\prime}\by_t$, and $R_4R_5R_6=27$ predictor factors, $\tilde{\bB}^{\prime}\by_{t-1}$, yielding a total of 54 latent factor time series.

 Although the model is estimated on disaggregated country- and commodity-level data, most dimension-reduced predictor and response factors closely track aggregated real-world global indicators. Because each column of $\bB_i$ is identified only up to sign, the sign of each factor is indeterminate; we orient each factor by choosing the sign that maximizes the correlation with the corresponding index.

Specifically, nearly all of the 54 factor series align with one of the three-month moving averages of standardized year-over-year growth rates for several key global indicators considered in the previous literature \citep{kilian2009not,caldara2022measuring,guichard2011dynamic,gopinath2015international}. These indicators include: (1) the Real Broad Dollar (RBD) index, (2) the Brent Crude Oil (BCO) price index, (3) the Geopolitical Risk (GPR) index, (4) the Index of Global Real Economic Activity (IGREA), (5) the U.S. Trade Policy Uncertainty (TPU) index, (6) the World Industrial Production (WIP) index, and (7) the World Semiconductor Trade Statistics (WSTS) index. \footnote{The matching procedure is based on the maximum correlation between each estimated factor series and the candidate indicator series. Specifically, each factor is assigned to the indicator that yields the highest correlation.}

\begin{figure}[H]
\centering
\includegraphics[width=1\textwidth]{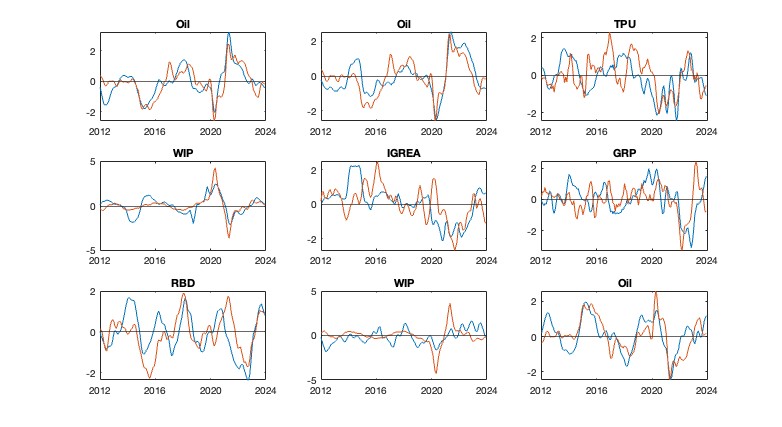} 
\caption{Response Factors for $r_3=1$ and Matched Economic Indicators }
\label{resp_fact_ld_1}
\end{figure}

Figure \ref{resp_fact_ld_1} presents the 3×3 standardized response factors for the three import and three export modes associated with the first commodity category, comparing the model-implied factors (red) with the corresponding observed indices (blue). The results for the remaining commodity categories (Figures \ref{resp_fact_ld_2}–\ref{resp_fact_ld_3}) are reported in Appendix D. Overall, the BCO price index is the most frequently matched indicator, corresponding to 30\% of the response factors. The RBD and IGREA indices are each associated with 15\% of the factors, followed by the GPR, WIP, and WSTS indices, each representing 11\% of the matched factors. The U.S. TPU index is the least frequently matched indicator, with a share of 7\%.

\begin{figure}[H]
\centering
\includegraphics[width=1\textwidth]{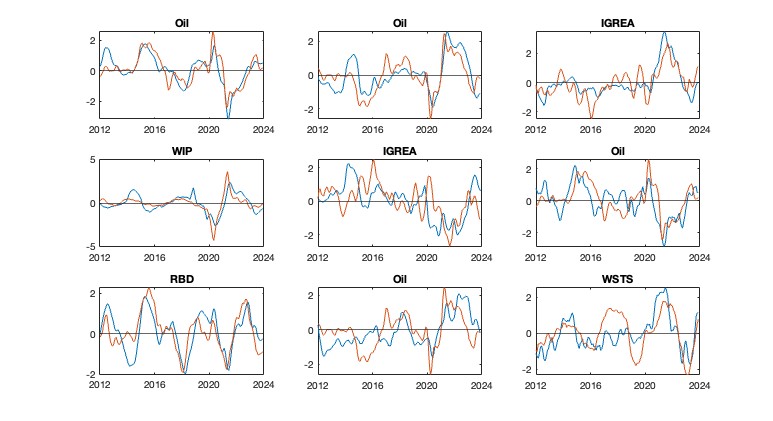} 
\caption{Predictor Factors for $r_3=1$ and Matched Economic Indicators}
\label{pred_fact_ld_1}
\end{figure}

Figure \ref{pred_fact_ld_1} shows the 3×3 standardized predictor factors for the first commodity category, with the results for the remaining two commodity categories also reported in Appendix D.    The two most frequently matched indicators, the RBD index and the BCO price index, account for approximately 50\% of the predictor factors, with shares of 30\% and 22\%, respectively. These are followed by the TPU index (15\%), the WIP index (15\%), and the IGREA index (11\%). The remaining indicators, including the WSTS and GPR indices, account for a relatively small fraction of the matched predictor factors, with shares of 7\% and 0\%, respectively.

 Combining the response and predictor factors, the two most frequently identified indicators are the Brent Crude Oil price index and the Real Broad Dollar index. This finding is consistent with the existing literature, which highlights the importance of these two indicators in capturing global economic conditions and international transmission mechanisms. Brent Crude Oil prices reflect fluctuations in global demand, supply conditions, and commodity market shocks, while the Real Broad Dollar index captures changes in global financial conditions, exchange rate movements, and the external value of the U.S. dollar. The strong alignment between the estimated latent factors and these indicators suggests that the proposed tensor autoregressive framework effectively extracts economically meaningful sources of variation underlying international trade dynamics. In addition, the identification of other indicators, including WIP, IGREA, TPU indices, highlights the role of both macroeconomic and uncertainty-related factors in shaping the evolution of global trade patterns.

\section{Conclusion}

In summary, this paper develops a flexible and computationally efficient Bayesian framework for third-order tensor autoregressions based on the Tucker decomposition. The proposed methodology incorporates stochastic volatility and global--local shrinkage priors into high-dimensional tensor-on-tensor regressions while extending the commonly used CP decomposition to the more general Tucker framework. Extensive Monte Carlo experiments show that the proposed Tucker models consistently outperform CP-based approaches, regardless of whether the true coefficient tensor is low rank, and also outperform conventional Bayesian VARs when the data-generating process is low rank. Moreover, Tucker models equipped with multiway stick-breaking shrinkage priors generally achieve better estimation accuracy, highlighting the benefits of adaptive regularization in high-dimensional settings. In the empirical application to an \(18\times 18\times 15\) tensor of bilateral trade flows across multiple commodity groups, the proposed framework identifies interpretable latent factors governing importer, exporter, and commodity interactions. These factors reveal the dominant roles of major trading economies, distinguish heterogeneous trade patterns across countries and commodity groups, and exhibit strong associations with key global macroeconomic indicators, including the Brent Crude Oil price index and the Real Broad Dollar index, providing an economically meaningful characterization of global trade dynamics.

\newpage

\section*{Appendix A: Proofs of Propositions}

\begin{proof}[Proof of Proposition \ref{B123}]
It suffices to establish the result for $i=1$, since the remaining cases follow analogously. We first state the following lemma.
\begin{lemma}
As noted in Section 3 of \citet{kolda2009tensor},  the mode$-i$ matricization of $\mB$ satisfies the following equation:
\begin{equation}\label{Bmat}
\begin{aligned}
 \mathbf{B}_{(i)} = \bB_i \bG_{(i)} \bB'_{-i}
\end{aligned}
\end{equation}
where \(\bG_{(i)}\) denotes the mode-\(i\) matricization of \(\mG\), and $\bB_{-i}
=
\displaystyle \bigotimes_{\substack{j=6 \\ j\ne i}}^{1} \bB_j$.
\end{lemma}

Without loss of generality, in the proof, we ignore the intercept term from the equations. According to Equation \eqref{Bmat}, to prove the proposition, it suffices to show that $$ \bY_{t,(1)} = \bB_{(1)} \bX_{1t} + \bE_{t,(1)} $$ Due to the redundancy of the subscript in $\bY_{t,(1)}$, We will represent $\bY_{t,(1)}$ as $\bY^{(1)}_t$ in this proof.

Let's start with the $(i_1, k)^{th}$ element of $\bY_{t,(1)}$ on the left side: $\bY^{(1)}_{t,i_1,k}$. For any $1\leq k\leq I_2I_3$, we can find a pair $(i_2,i_3)$ such that $k=(i_3-1)I_2+i_2$. Specifically, $i_3=1+\lceil (k-1)/I_2\rceil$, $i_2=1+\bmod (k-1, I_2).$    By definition of mode-$1$ matricization,  $\bY^{(1)}_{t,i_1,k}$ corresponds to the $(i_1,i_2,i_3)^{th}$ element of $\mY_t$. Thus, $\bY^{(1)}_{t,i_1,k}=\mY_{t,i_1,i_2,i_3}$. Likewise, $(\bE_{t,(1)})_{i_1,k}=\mE_{t,i_1,i_2,i_3}$.

As for the right-hand side, $\bX_{1t}= \by_{t-1} \otimes \bI_{I_2I_3}$, is an $I_2I_3I\times I_2I_3$ matrix. The  $(i_1, k)^{th}$ element of $\bB_{(1)} \bX_{1t}$ is the sum of the products of the elements in the $i_1^{th}$ row of $\bB_{(1)}$ and the $k^{th}$ column of $\bX_{1t}$.  The $j^{th}$ element in the $k^{th}$ column of $\bX_{1t}$ is:
\[
\begin{aligned}
    \bX_{1t,j,k} &= \begin{cases} \by_{t-1,m}, & \text{if } j=k+(m-1)I_2I_3 \text{ for } m=1,2,\ldots,I, \\ 0, & \text{otherwise.} \end{cases}
\end{aligned}
\]

Thus the  $(i_1, k)^{th}$ element of $\bB_{(1)} \bX_{1t}$ has the following representation:

$$
\begin{aligned}
\sum_{j=1}^{I_2I_3I}\bB_{(1),i_1,j}\bX_{1t,j,k}=&\sum_{m=1}^{I}\bB_{(1),i_1,k+(m-1)I_2I_3}\bX_{1t,k+(m-1)I_2I_3,k}\\
=&\sum_{m=1}^{I}\bB_{(1),i_1,(m-1)I_2I_3+(i_3-1)I_2+i_2}\by_{t-1,m}
\end{aligned}
$$
Note that for any $1\leq m\leq I$, there exists a triplet of $(i_4,i_5,i_6)$ such that $m=(i_6-1)I_1I_2+(i_5-1)I_1+i_4$. Specifically,  $i_6=1+\lceil (m-1)/  I_1I_2\rceil$, $i_5=1+\lceil \bmod (m-1, I_1I_2) / I_1\rceil$ and  $i_4=1+ \bmod (m-1, I_1)$.
Thus, $\bB_{(1), i_1, (m-1) I_2 I_3 + (i_3 - 1) I_2 + i_2}$ is equivalent to $\bB_{(1), i_1, 1 + \sum_{u=2}^6 (i_u - 1) J_u}$, where $J_u = \prod_{v=2}^{u-1} I_v$, with $I_4 = I_1$, $I_5 = I_2$, and $I_6 = I_3$, which corresponds to the $(i_1,i_2,i_3,i_4,i_5,i_6)^{th}$ element of $\mB$. Then the above equation can be rewritten as 
$$
\begin{aligned}
\sum_{j=1}^{I_2I_3I}\bB_{(1),i_1,j}\bX_{1t,j,k}&=\sum_{i_4=1}^{I_1}\sum_{i_5=1}^{I_2}\sum_{i_6=1}^{I_3}\mB_{i_1,i_2,i_3,i_4,i_5,i_6}\by_{t-1,(i_6-1)I_1I_2+(i_5-1)I_1+i_4}
\\&=\sum_{i_4=1}^{I_1}\sum_{i_5=1}^{I_2}\sum_{i_6=1}^{I_3}\mB_{i_1,i_2,i_3,i_4,i_5,i_6}\mY_{t-1,i_4,i_5,i_6}
\end{aligned}
$$
It follows that $$\left(\bB_{(1)} \bX_{1t}+\bE_{t,(1)}\right)_{i_1,k}=\sum_{i_4=1}^{I_1}\sum_{i_5=1}^{I_2}\sum_{i_6=1}^{I_3}\mB_{i_1,i_2,i_3,i_4,i_5,i_6}\mY_{t-1,i_4,i_5,i_6}+\mE_{t,i_1,i_2,i_3}=\mY_{t,i_1,i_2,i_3}$$
This demonstrates the equality of the left and right sides of the equation \eqref{TARtoMAR} for $i=1$.
\end{proof}

\begin{proof}[Proof of Proposition \ref{MVAR}]
According to equation \eqref{TARtoMAR} for $i=1$, we have
$$ \bY_{t,(1)}-\bA_{t,(1)}=\bB_1\bG_{(1)}\bB_{-1}'(\by_{t-1}\otimes \bI_{I_2I_3})+ \bE_{t,(1)}$$
 Upon vectorizing both sides, we have 
\begin{equation*}\label{B456}
\begin{aligned}
  \by_t-\ba_t=&\tvec(\bB_1\bG_{(1)}\bB_{-1}'(\by_{t-1}\otimes \bI_{I_2I_3}))+ \be_t\\=& \left((\by_{t-1}'\otimes \bI_{I_3}\otimes \bI_{I_2})\bB_{-1} \otimes \bB_1\right)\tvec(\bG_{(1)})+\be_t \\=&  (\by_{t-1}'\tilde{\bB}\otimes\bB_3\otimes\bB_2 \otimes \bB_1)\bg+\be_t \\=& (\by_{t-1}'\tilde{\bB}\otimes\tilde{\bB}_m)\tvec(\tilde{\bG})+\be_t\\=&\tilde{\bB}_m\tilde{\bG}{\tilde{\bB}'\by_{t-1}}+\be_t
\end{aligned}
\end{equation*} 
where $\tilde{\bB}=\bB_6\otimes\bB_5\otimes\bB_4$, $\tilde{\bB}_m=\bB_3\otimes\bB_2\otimes\bB_1$ and $\tilde{\bG}$ is an $\tilde{R}_m\times \tilde{R}$ matrix satisfying $\tvec(\tilde{\bG})=\tvec(\mG)$, where $\tilde{R}_m=R_1R_2R_3$ and $\tilde{R}=R_4R_5R_6$.

\end{proof}

\section*{Appendix B: Additional Estimation Details} 

\subsection*{B.1 Full conditional distribution of the global-local shrinkage parameters }

The sampler for the shrinkage parameters are given as follows.

 1. For each global parameter $\tau_i$,
 

$(\tau_i\mid \bB_i, \{\phi_r^{{i}}\}_{r=1}^{R_i}) \sim \text{GIG}\left(\alpha_{\tau_i}-\frac{R_iI_i}{2},2\beta_{\tau_i},\sum_{{r}=1}^{R_i}\bfb_{r}^{(i)'}(\phi_r^{i})^ {-1}\bfb_r^{(i)} \right)$

2. For each auxiliary variable $\eta_r^{i}$, if $r < R_i$,
 
$\begin{aligned}
p(\eta_r^{i}&\mid \bB_i,\tau_i,\alpha_i)\propto p(\eta_r^{i}\mid \alpha_i)\prod_{u={r}}^{R_i} p(\bfb_u^{(i)}\mid \eta_r^{i},\tau_i) \propto (1-\eta_r^{i})^{\alpha_i-1}\prod_{u={r}}^{R_i}\phi_u^{-\frac{I_i}{2}} e^{-\frac{\bfb_u^{{(i)}'}\bfb_u^{(i)}}{2\tau_i\phi_u}}\\ \propto &(1-\eta_r^{i})^{\alpha_i-1}\prod_{u={r}}^{R_i-1}\left(\eta_u^i\prod_{l=1}^{u-1}(1-\eta_l^i)\right)^{-\frac{I_i}{2}}\left(\prod_{l=1}^{R_i-1}(1-\eta_l^i)\right)^{-\frac{I_i}{2}}
e^{-\frac{1}{2}\left(\frac{\bfb_u^{{(i)}'}\bfb_{u}^{(i)}}{\tau_i\eta_u^i\prod_{l=1}^{u-1}(1-\eta_l^i)} +\frac{\bfb_{R_i}^{(i)'}\bfb_{R_i}^{(i)}}{\tau_i\prod_{l=1}^{R_i}(1-\eta_l^i)}\right)}
\\   \propto &(1-\eta_r^{i})^{\alpha_i-\frac{(R_i-{r})I_i}{2} -1}(\eta_r^{i})^{-\frac{ I_i }{2}}e^{- \frac{1}{2\tau_i}\left(\sum_{u={r}}^{R_i-1}\frac{\bfb_u^{{(i)}'}\bfb_{u}^{(i)}}{\eta_u^i \prod_{l=1}^{u-1}\left(1-\eta_l^i\right)}+\frac{\bfb_{R_i}^{(i)'}\bfb_{R_i}^{(i)}}{\prod_{l=1}^{R_i-1}\left(1-\eta_l^i\right)}\right)}
\end{aligned}$

Since the posterior distribution is not available in closed form, we need to use the Metropolis-Hastings (MH) algorithm to sample $\eta$. Following \citet{guhaniyogi2021bayesian}, we apply the random-walk MH algorithm with a normal distribution having a variance of $0.01^2$.   After drawing $\eta$, set $\phi_r^{i}=\eta_r^{i} \prod_{l=1}^{{r}-1}\left(1-\eta_l^i\right)$, and $\phi_R^i=$ $1-\sum_{{r}=1}^{R-1} \phi_r^{i}$.

3. For each auxiliary variable $\alpha_i$,

Let the prior distribution of $\alpha_i$ be uniform. The posterior distribution of $\alpha_i$ is then expressed as:$$p(\alpha_i\mid\eta_i)\propto\prod_{r=1}^{R_i}p(\eta_r^{i}\mid\alpha_i) \propto \alpha_i^{ R_i}\prod_{r=1}^{R_i}(1-\eta_r^{i})^{\alpha_i-1} $$


We use Griddy-Gibbs Sampler to sample the posterior distribution of $(\alpha_i\mid\eta_i)$. Specifically, we define a grid over the interval $[0,1]$ and compute the posterior density of $\alpha_i$ at each point on the grid as the weights to sample $\alpha_i$.



\subsection*{B.2 Full conditional distribution of $\bB_2$ and $\bB_3$ }


The conditional posterior distributions of $\bfb_2$ and $\bfb_3$ are obtained in a similar manner as that of $\bfb_1$. The details are omitted for brevity.

The postetior distribution of $(\bfb_2 \mid \mY, \{\bB_i\}_{i\ne 2}, \bG_2, \tilde{\vSigma}_2, \vomega)$ is also a Gaussian distribution:

\begin{equation}\label{B_2}(\bfb_2\mid\mY, \{\bB_i\}_{i\ne 2},\mG, \tilde{\vSigma}_2, \vomega ) \sim \distn{N} (\widehat{\mathbf{b}}_2,\mathbf{K}_{\mathbf{b}_2}^{-1}) \end{equation}
where 
$$
\begin{aligned}\bK_{\bfb_2}=&\bV_{\bfb_2}^{-1}+\sum_{t=1}^T \omega_t^{-1}\left(\bG_{(2)}\bB_{-2}^{\prime} \bX_{2t}(\vSigma_3^{-1}\otimes \vSigma_1^{-1}) \bX_{2t}^{\prime}\bB_{-2}\bG_{(2)}'\right)\otimes \vSigma_2^{-1} \\ \hat{\bfb}_2=&\bK_{\bfb_2}^{-1}\left(\bV_{\bfb_2}^{-1} \bfb_{20}+\sum_{t=1}^T \tvec( \omega_t^{-1} \vSigma_2^{-1}(\bY_{t, (2)}-\bA_{t,(2)})(\vSigma_3^{-1}\otimes \vSigma_1^{-1})\bX_{2t}'\bB_{-2}\bG'_{(2)} )\right)\end{aligned}$$

The posterior distribution of $\bfb_3$ is similarly given by
\begin{equation}\label{B_3}(\bfb_3\mid\mY,\{\bB_i\}_{i\ne 3},\mG, \tilde{\vSigma}_3, \vomega) \sim \distn{N} (\widehat{\mathbf{b}}_3,\mathbf{K}_{\mathbf{b}_3}^{-1}) \end{equation}
where 
$$
\begin{aligned}\bK_{\bfb_3}=&\bV_{\bfb_3}^{-1}+\sum_{t=1}^T \omega_t^{-1}\left(\bG_{(3)}\bB_{-3}^{\prime} \bX_{3t}(\vSigma_2^{-1}\otimes \vSigma_1^{-1}) \bX_{3t}^{\prime}\bB_{-3}\bG_{(3)}'\right)\otimes \vSigma_3^{-1} \\ \hat{\bfb}_3=&\bK_{\bfb_3}^{-1}\left(\bV_{\bfb_3}^{-1} \bfb_{30}+\sum_{t=1}^T \tvec( \omega_t^{-1} \vSigma_3^{-1}(\bY_{t, (3)}-\bA_{t,(3)})(\vSigma_2^{-1}\otimes \vSigma_1^{-1})\bX_{3t}'\bB_{-3}\bG'_{(3)} )\right)\end{aligned}$$

\subsection*{B.3 Full conditional distribution of $\bB_5$ and $\bB_6$ }

In this appendix, we derive the posterior distributions for $\bB_5$ and $\bB_6$. We begin with a  useful proposition.

\begin{prop}\label{BB5}
\begin{equation}
   \tilde{\bB}'\by_{t-1}=(\bP'(\bB'_5\otimes\bB'_6)\otimes \bB'_4)\tvec(\bY_{t-1,(2)}')
\end{equation}
where $\bP$ is an $R_5R_6\times R_5R_6 $ commutation matrix so that $\mathbf{P} \operatorname{vec}(\mathbf{Z})=\operatorname{vec}\left(\mathbf{Z}^{\prime}\right)$ for any $R_5 \times R_6$ matrix $\mathbf{Z}$. Specifically, $\bP_{k,q}=1$ if there exist $r_5$ and $r_6$ such that $k=(r_5-1)R_6+r_6$ and $q=(r_6-1)R_5+r_5$; otherwise, $\bP_{k,q}=0$.
\end{prop}

\begin{proof} 
Consider the $j^{th}$ element on both sides. For any $1\leq j\leq \tilde{R}$, there exists a triplet of $(j_1,j_2,j_3)$ such that $j=(j_3-1)R_4R_5+(j_2-1)R_4+j_1$. Specifically,  $j_3=1+\lceil (j-1)/  R_4R_5\rceil$, $j_2=1+\lceil \bmod (j-1, R_4R_5) / R_4\rceil$ and  $j_1=1+ \bmod (j-1, R_4)$. Similarly, for any $1\leq i\leq I$, there exists a unique set of $(i_1,i_2,i_3)$ such that $i=(i_3-1)I_1I_2+(i_2-1)I_1+i_1$. Then the $j^{th}$ element on the left-hand side is given by 
\begin{equation*}
\begin{aligned}
\sum_{i=1}^{I}&(\tilde{\bB}')_{j,i}\by_{t-1,i}=\sum_{i=1}^{I}\tilde{\bB}_{i,j}\by_{t-1,i}\\&=
\sum_{i_1=1}^{I_1}\sum_{i_2=1}^{I_2}\sum_{i_3=1}^{I_3}\tilde{\bB}_{(i_3-1)I_1I_2+(i_2-1)I_1+i_1,(j_3-1)R_4R_5+(j_2-1)R_4+j_1} \\ & \times \by_{t-1,(i_3-1)I_1I_2+(i_2-1)I_1+i_1} \\&=
\sum_{i_1=1}^{I_1}\sum_{i_2=1}^{I_2}\sum_{i_3=1}^{I_3}\bB_{4,i_1,j_1}\bB_{5,i_2,j_2}\bB_{6,i_3,j_3}\mY_{t-1,i_1,i_2,i_3} 
   \end{aligned}
\end{equation*}

Likewise, for any $1 \leq m \leq I_2 I_3$, $1 \leq o \leq R_5 R_6$, and $1 \leq o \leq R_5 R_6$, there exist pairs $(m_2, m_3)$, $(l_2, l_3)$ and $(o_2, o_3)$ such that
$m = (m_2 - 1) I_3 + m_3, l = (l_3 - 1) R_5 + l_2, \text{ and } o = (o_2 - 1) R_6 + o_3.$
Then the $(m,l)^{th}$ element of $(\bB_5 \otimes \bB_6)\bP$ on the right-hand side is given by
\begin{equation}\label{BBP}
\begin{aligned}
\sum_{o=1}^{R_5R_6}(\bB_5\otimes\bB_6)_{m,o}\bP_{o,l}=&\sum_{o_2=1}^{R_5}\sum_{o_3=1}^{R_6}(\bB_5\otimes\bB_6)_{(m_2-1)I_3+m_3,(o_2-1)R_6+o_3}\bP_{(o_2-1)R_6+o_3,l} \\=&\sum_{o_2=1}^{R_5}\sum_{o_3=1}^{R_6}\bB_{5,m_2,o_2}\bB_{6,m_3,o_3}\bP_{(o_2-1)R_6+o_3,(l_3-1)R_5+l_2}= \bB_{5,m_2,l_2}\bB_{6,m_3,l_3}
\end{aligned}
\end{equation}
The last equality holds due to the definition of $\bP$. 

In addition, $\mY_{t-1,i_1,i_2,i_3}$ corresponds to the $(i_2, (i_3 - 1) I_1 + i_1)$ element of $\bY_{t-1,(2)}$, and hence to the $((i_3 - 1) I_1 + i_1, i_2)$ element of $\bY_{t-1,(2)}'$, or equivalently, to the $i = (i_2 - 1) I_1 I_3 + (i_3 - 1) I_1 + i_1$ element of $\tvec(\bY_{t-1,(2)}')$.

Accordingly, the $j^{th}$ element of the right side can be represented as follows: 
\begin{equation*}
\begin{aligned}
\sum_{i=1}^I&(\bP'(\bB'_5\otimes\bB'_6)\otimes \bB'_4)_{j,i}\tvec(\bY_{t-1,(2)}')_i=\sum_{i=1}^I((\bB_5\otimes\bB_6)\bP) \otimes \bB_4)_{i,j}\tvec(\bY_{t-1,(2)}')_i\\=&\sum_{i_1=1}^{I_1}\sum_{i_2=1}^{I_2}\sum_{i_3=1}^{I_3}((\bB_5\otimes\bB_6)\bP) \otimes \bB_4)_{(i_2-1)I_1I_3+(i_3-1)I_1+i_1,(j_3-1)R_4R_5+(j_2-1)R_4+j_1} \\ & \times \tvec(\bY_{t-1,(2)}')_{(i_2-1)I_1I_3+(i_3-1)I_1+i_1}\\=&\sum_{i_1=1}^{I_1}\sum_{i_2=1}^{I_2}\sum_{i_3=1}^{I_3}((\bB_5\otimes\bB_6)\bP)_{(i_2-1)I_3+i_3,(j_3-1)R_5+j_2} \bB_{4,i_1,j_1}\mY_{t-1,i_1,i_2,i_3}\\=&\sum_{i_1=1}^{I_1}\sum_{i_2=1}^{I_2}\sum_{i_3=1}^{I_3}(\bB_{5,i_2,j_2} \bB_{6,i_3,j_3}) \bB_{4,i_1,j_1}\mY_{t-1,i_1,i_2,i_3}
\end{aligned}
\end{equation*}
The last equality follows the equation $\eqref{BBP}$, thus establishing equivalence between the left and right sides. Proof concluded.
\end{proof}

Based on Proposition $ \ref{BB5} $, we can separate $ \bB_5 $ from the remaining Tucker components. It follows that
\begin{equation*}
\begin{aligned}
    \tilde{\bB}'\by_{t-1}&=(\bP'(\bB'_5\otimes\bB'_6)\otimes \bB'_4)\tvec(\bY_{t-1,(2)}')\\&=(\bP'\otimes \bI_{R_4})(\bB'_5\otimes\bB'_6\otimes\bB'_4)\tvec(\bY_{t-1,(2)}')\\&=(\bP'\otimes \bI_{R_4})\tvec\left(((\bB'_6\otimes\bB'_4)\bY_{t-1,(2)}'\bB_5)\right)\\&=(\bP'\otimes \bI_{R_4})\left(\bI_{R_5}\otimes(\bB'_6\otimes\bB'_4)\bY_{t-1,(2)}'\right)\tvec(\bB_5)\\&=(\bP'\otimes \bI_{R_4})\left(\bI_{R_5}\otimes\bB'_6\otimes\bB'_4\right)(\bI_{R_5}\otimes\bY_{t-1,(2)}')\tvec(\bB_5)
    \\&=(\bP'(\bI_{R_5}\otimes\bB'_6)\otimes \bB'_4)(\bI_{R_5}\otimes\bY_{t-1,(2)}')\tvec(\bB_5)
    \end{aligned}
\end{equation*}
Then we have
\begin{equation*}
\begin{aligned}
  \by_t= \tilde{\bB}_m\tilde{\bG}(\bP'(\bI_{R_5}\otimes\bB'_6)\otimes \bB'_4)(\bI_{R_5}\otimes\bY_{t-1,(2)}')\tvec(\bB_5)+\be_t
\end{aligned}
\end{equation*} 
Let $\bfb_5=\operatorname{vec}(\bB_5)$, and assume the Gaussian prior $\bfb_5  \sim \distn{N}(\bfb_{50},\bV_{\bfb_5})$,
where $\bfb_{50}$ is a vector of length $I_2R_5$, and $\bV_{\bfb_5}$ is an $I_2R_5\times I_2R_5$ diagonal matrix. Then the posterior distribution of $\bfb_5$ is given by:
$$(\bfb_5|\mY, \{\bB_i\}_{i\ne 5},{\mG}, \vSigma, \vomega) \sim \distn{N}(\hat{\bfb}_5,\bK^{-1}_{\bfb_5})$$
where $$
\begin{aligned}
\bK_{\bfb_5}=&\bV_{\bfb_5}^{-1}+\sum_{t=1}^T\omega_t^{-1}(\bI_{R_5}\otimes\bY_{t-1,(2)})((\bI_{R_5}\otimes\bB_6)\bP\otimes \bB_4)\tilde{\bG}'\tilde{\bB}'_m\vSigma^{-1}\tilde{\bB}_m\tilde{\bG} \\ & \times (\bP'(\bI_{R_5}\otimes\bB'_6)\otimes \bB'_4)(\bI_{R_5}\otimes\bY_{t-1,(2)}')\end{aligned}$$
$$\hat{\bfb}_5=\bK_{\bfb_5}^{-1}\left(\bV^{-1}_{\bfb_5}\bfb_{50}+\sum_{t_1}^T\omega_t^{-1}(\bI_{R_5}\otimes\bY_{t-1,(2)})((\bI_{R_5}\otimes\bB_6)\bP\otimes \bB_4)\tilde{\bG}'\tilde{\bB}'_m\vSigma^{-1}\by_t\right)$$
Finally, let $\bfb_6=\tvec(\bB_6)$. Then equation \eqref{eq:MVAR} can be rewritten as:
\begin{equation}\label{eq:B6}
\by_t=\ba_t+\tilde{\bB}_m\tilde{\bG}\left(\bI_{R_6} \otimes (\bB'_5\otimes\bB'_4)\bY_{t-1,(3)}'\right)\bfb_6+\be_t
\end{equation} 
Suppose the prior on $ \bfb_6 $ is given by
$ \bfb_6  \sim \distn{N}(\bfb_{60},\bV_{\bfb_6})$,
where $\bfb_{60}$ is an $I_3R_6$ dimensional vector, and $\bV_{\bfb_6}$ is an $I_3R_6\times I_3R_6$ diagonal matrix. Then the posterior distribution of $\bfb_6$ is as follows:
$$(\bfb_6|\mY, \{\bB_i\}_{i\ne 6},{\mG}, \vSigma, \vomega) \sim \distn{N}(\hat{\bfb}_6,\bK^{-1}_{\bfb_6})$$
where 
\begin{equation*}
\begin{aligned}
\bK_{\bfb_6}=&\bV_{\bfb_6}^{-1}+\sum_{t=1}^T\omega_t^{-1}\left(\bI_{R_6}\otimes\bY_{t-1,(3)}(\bB_5\otimes\bB_4)\right)\tilde{\bG}'\tilde{\bB}'_m\vSigma^{-1}\tilde{\bB}_m\tilde{\bG}\left(\bI_{R_6}\otimes(\bB'_5\otimes\bB'_4)\bY_{t-1,(3)}'\right)\\ \hat{\bfb}_6=&\bK_{\bfb_6}^{-1}\left(\bV^{-1}_{\bfb_6}\bfb_{60}+\sum_{t=1}^T\omega_t^{-1}\left(\bI_{R_6}\otimes\bY_{t-1,(3)}(\bB_5\otimes\bB_4)\right)\tilde{\bG}'\tilde{\bB}'_m\vSigma^{-1}\by_t\right)\end{aligned}\end{equation*}

\section*{Appendix C: Data Description}\label{datadescrip}
The fifteen commodity groups include all two-digit HS categories: Animal and Animal Products (HS01–HS05), Vegetable Products (HS06–HS15), Foodstuffs (HS16-HS24), Mineral Products (HS25–HS27), Chemicals and Allied Industries (HS28–HS38), Plastics and Rubbers (HS39, HS40), Raw Hides, Skins, Leather, and Furs (HS41–HS43), Wood and Wood Products (HS44–HS49), Textiles (HS50–HS63), Footwear and Headgear (HS64–HS67), Stone and Glass (HS68–HS71), Metals (HS72–HS83), Machinery and Electrical (HS84, HS85), Transportation (HS86–HS89), and Miscellaneous  (HS90–HS97).

\section*{Appendix D: Additional Response and Predictor Factors}\label{factorfigs}

This appendix reports the remaining response and predictor factor time series panels.




\begin{figure}[H]
\centering
\includegraphics[width=0.8\textwidth]{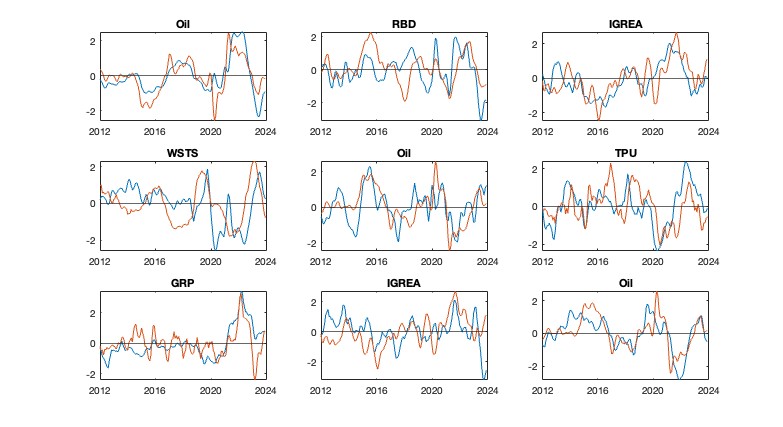} 
\caption{Response Factors for $r_3=2$ and Matched Economic Indicators }
\label{resp_fact_ld_2}
\end{figure}

\begin{figure}[H]
\centering
\includegraphics[width=0.8\textwidth]{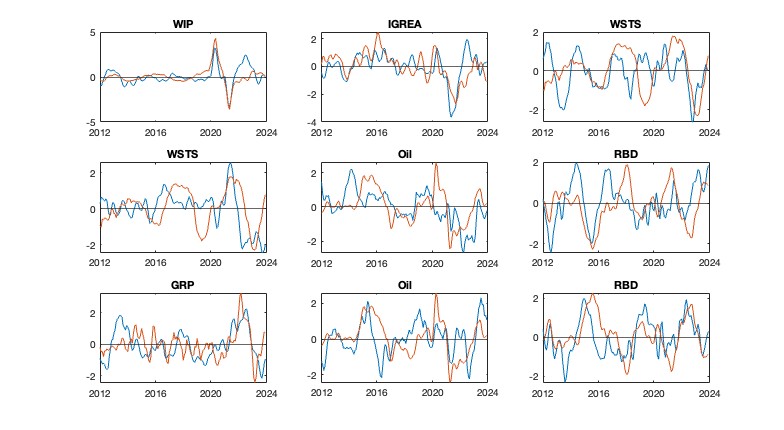} 
\caption{Response Factors for $r_3=3$ and Matched Economic Indicators }
\label{resp_fact_ld_3}
\end{figure}

\begin{figure}[H]
\centering
\includegraphics[width=0.8\textwidth]{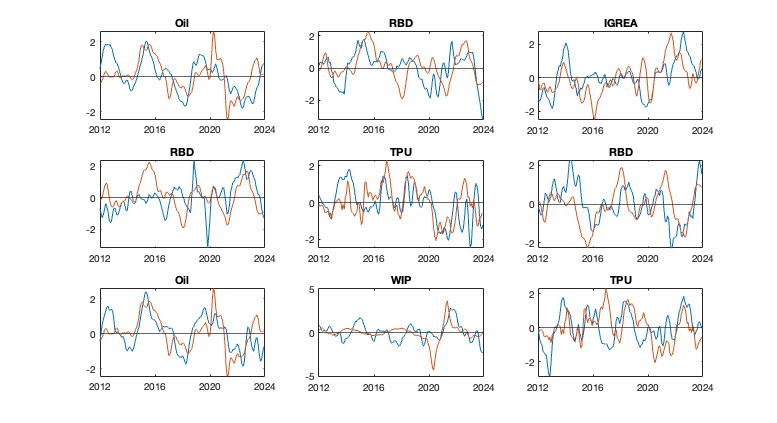} 
\caption{Predictor Factors for $r_3=2$ and Matched Economic Indicators }
\label{pred_fact_ld_2}
\end{figure}

\begin{figure}[H]
\centering
\includegraphics[width=0.8\textwidth]{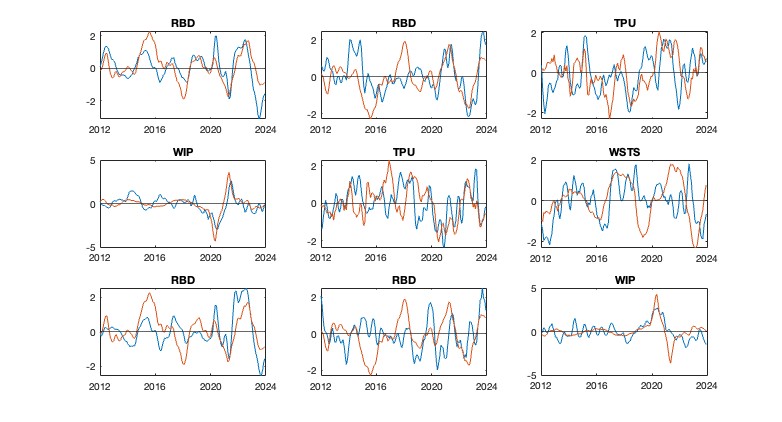} 
\caption{Predictor Factors for $r_3=3$ and Matched Economic Indicators }
\label{pred_fact_ld_3}
\end{figure}

\newpage
\singlespace
\bibliographystyle{econometrica}
\bibliography{BTAR}

\end{document}